\documentclass[
aps
,prx
,twocolumn,10pt
,superscriptaddress
,longbibliography
,amsmath
,amssymb
,amsfonts
]{revtex4-1}
\usepackage{float}
\usepackage{commath}
\usepackage{graphicx}
\usepackage{verbatim}
\usepackage{color}
\usepackage[dvipsnames,svgnames,x11names,hyperref]{xcolor}

\usepackage{times}
\usepackage[compat=1.1.0]{tikz-feynman}
\usepackage[sf,tight]{subfigure}
\usepackage[toc,page]{appendix}
\usepackage{siunitx}
\DeclareSIUnit\gauss{G}

\definecolor{linkcolor}{RGB}{6,69,173} 
\definecolor{diffcolor}{RGB}{175,31,36} 

\usepackage[colorlinks=true,
            linkcolor=linkcolor,
            urlcolor=linkcolor,
            citecolor=linkcolor,
            unicode,
            pdfencoding=auto]{hyperref}

\usepackage[
]{physpack}

\newcommand{\subfigimg}[3][,]{%
  \setbox1=\hbox{\includegraphics[#1]{#3}}
  \leavevmode\rlap{\usebox1}
  \rlap{\hspace*{9pt}\raisebox{\dimexpr\ht1-1\baselineskip}{#2}}
  \phantom{\usebox1}
}

\newcommand{\COMMENT}[1]{}

\begin{document}

\title{Triangular lattice Hubbard model physics at intermediate temperatures}

\author{Kyungmin Lee}
\affiliation{Department of Physics, Florida State University, Tallahassee, Florida 32306, USA}
\affiliation{National High Magnetic Field Laboratory, Tallahassee, Florida 32310, USA}

\author{Prakash Sharma}
\affiliation{Department of Physics, Florida State University, Tallahassee, Florida 32306, USA}
\affiliation{National High Magnetic Field Laboratory, Tallahassee, Florida 32310, USA}

\author{Oskar Vafek}
\affiliation{Department of Physics, Florida State University, Tallahassee, Florida 32306, USA}
\affiliation{National High Magnetic Field Laboratory, Tallahassee, Florida 32310, USA}

\author{Hitesh J. Changlani}
\email{hchanglani@fsu.edu}
\affiliation{Department of Physics, Florida State University, Tallahassee, Florida 32306, USA}
\affiliation{National High Magnetic Field Laboratory, Tallahassee, Florida 32310, USA}

\begin{abstract}
Moire systems offer an exciting playground to study many-body effects of strongly correlated electrons in regimes that are not easily accessible in conventional material settings. 
Motivated by a recent experiment on $\text{WSe}_2/\text{WS}_2$ moire bilayers [Y. Tang et al., Nature 579, 353–358 (2020)], which realizes a triangular superlattice 
with a small hopping (of approximately 10 Kelvin), with tunable density of holes, we explore the Hubbard model on the triangular lattice 
for intermediate temperatures $t \lesssim T < U$. Employing finite temperature Lanczos calculations, and closely following the fitting protocols used in the experiment, 
we recover the observed trends in the reported Curie-Weiss temperature $\Theta$ with filling, using the reported interaction strength $U/t=20$. 
We focus on the large increase of $|\Theta|$ on decreasing the density below half filling and the sign change of $\Theta$ at higher fillings, which signals the onset of ferromagnetism. The increase in $|\Theta|$ is also seen in the $t$-$J$ model (the low energy limit of the Hubbard model) in the intermediate temperature range, which we clarify is opposite to the trend in its high temperature limit. 
Differences between the low, intermediate and high temperature regimes are discussed. Our numerical calculations also capture the crossover between short-range 
antiferromagnetic to ferromagnetic order in the intermediate temperature regime, a result broadly consistent with the experimental findings. We find that this behavior 
is a finite-temperature remnant of the underlying zero temperature phase transition, which we explore with ground state density matrix renormalization group calculations. 
We provide evidence of ferromagnetism characterized by weak (but robust) correlations that explain the small $\Theta$ seen in the experiment.
\end{abstract}

\maketitle

\section{Introduction} 
\label{sec:intro}
Magnetism in strongly correlated electronic systems poses fundamental questions related to the intricate ways 
electrons can order (or fail to order) in different settings and conditions - temperature, lattice geometry, frustrated interactions, 
spin-orbit coupling etc.~\cite{BalentsQSL}. While there has been tremendous progress in our understanding 
of low-temperature and ground state properties of such systems~(see for example,~\cite{Simons_Hubbard,Arovas_Hubbard,Qin_Hubbard,NormanRMP,FutureSC}), far less is definitively 
understood about their finite temperature properties and response. 
The ``intermediate temperature scale" (temperature larger than hopping but smaller than the interaction strength) in real materials can be rather large ($\sim 1000 $ K or more) and is hence difficult to access experimentally. 
This situation changed with a recent breakthrough in engineering moire systems~\cite{cao2018n,BM_Model,tang2020n,BalentsMoire,KangVafek} which realize a triangular superlattice with significantly renormalized parameters but with relative interaction to kinetic energy strengths (eg. $U/t$ in the Hubbard model) 
comparable to other strongly correlated materials, such as the high $T_c$ superconducting cuprates. Moire systems thus offer an exciting platform to study 
many-body effects of strongly correlated electrons in regimes that are not easily accessible in conventional material settings. 

\begin{figure}
\hspace{-15pt}\text{}
\subfigimg[height=137pt]
{\qquad\qquad\qquad\qquad\qquad\qquad\textsf{\normalsize(a)} \qquad\qquad\qquad\qquad \;\;\;\;\;\;\;\;\;\; \textsf{\normalsize(b)}}
{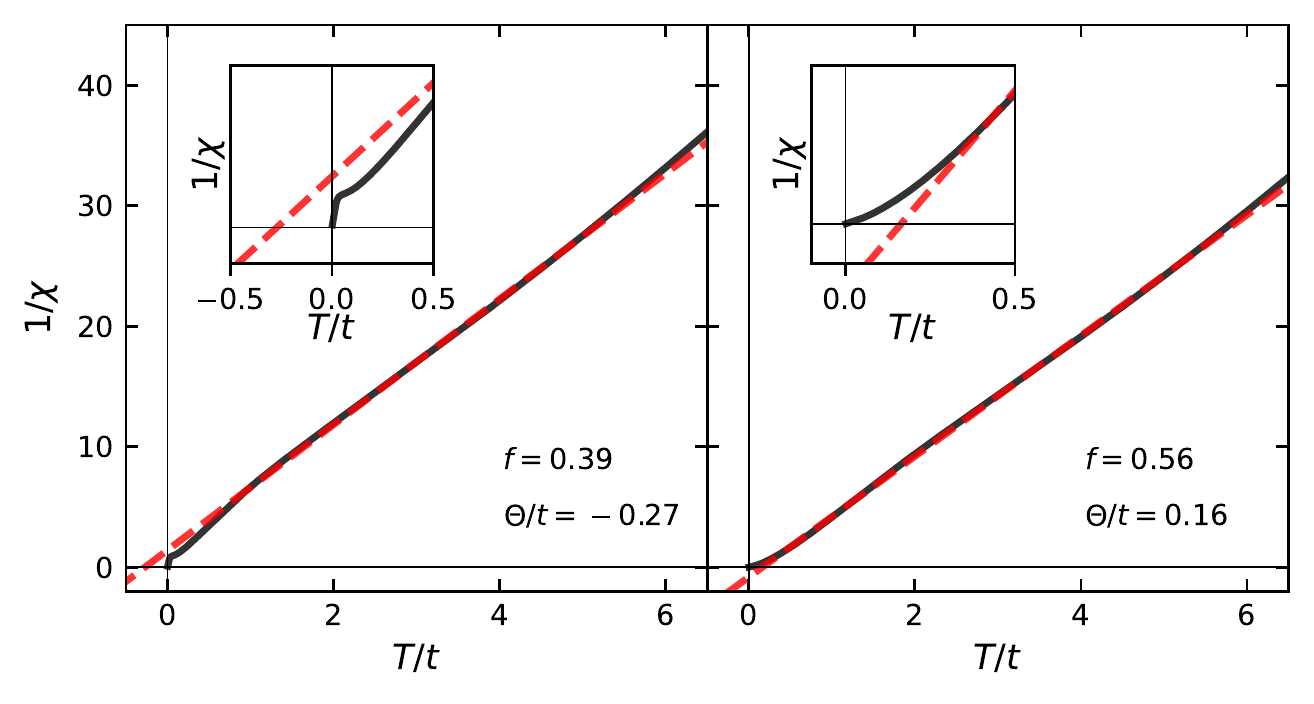}
\caption{
Inverse susceptibility ($1/\chi$) versus temperature ($T$, in units of $t$) for the T-9 cluster for fillings (a) $f=0.39$ and (b) $f=0.56$. Red dashed lines represent the linear fits to the data for $T/t \in [0.8, 5.5]$. Insets show the range where the linear fits intersect the horizontal temperature axis, the intercept yields the Curie-Weiss temperature $\Theta$. Similar analyses are performed for the T-12 and T-15 clusters with the finite temperature Lanczos method to generate Fig.~\ref{fig:window_CW}(a).}
\label{fig:expt_theory}
\end{figure}

\begin{figure*}
\subfigimg[height=135pt]{\qquad\qquad\;\;\textsf{\normalsize(a)}}{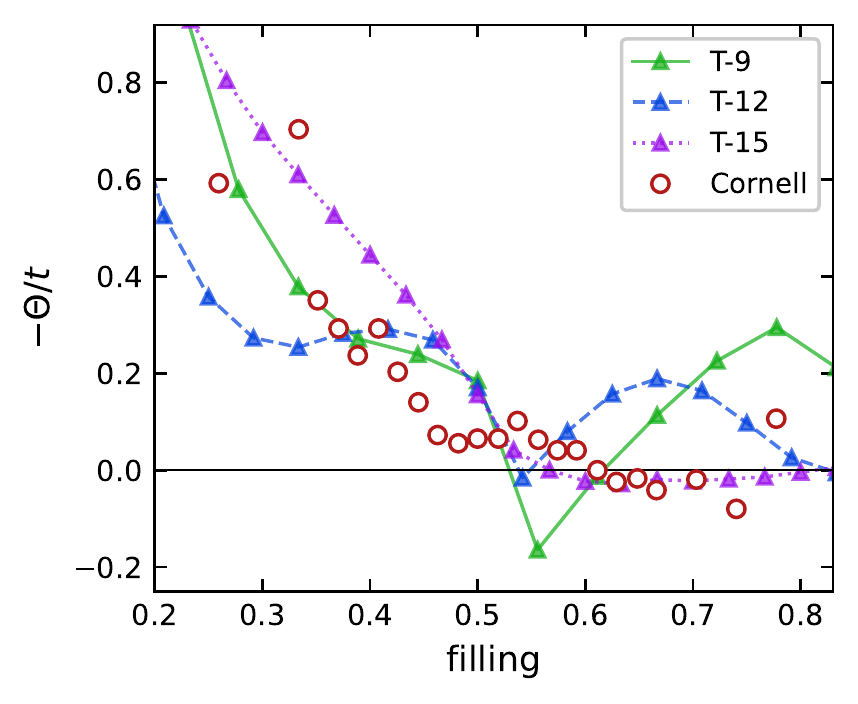}%
\subfigimg[height=132pt,width=170pt]{%
\qquad\textsf{\normalsize(b)}
}{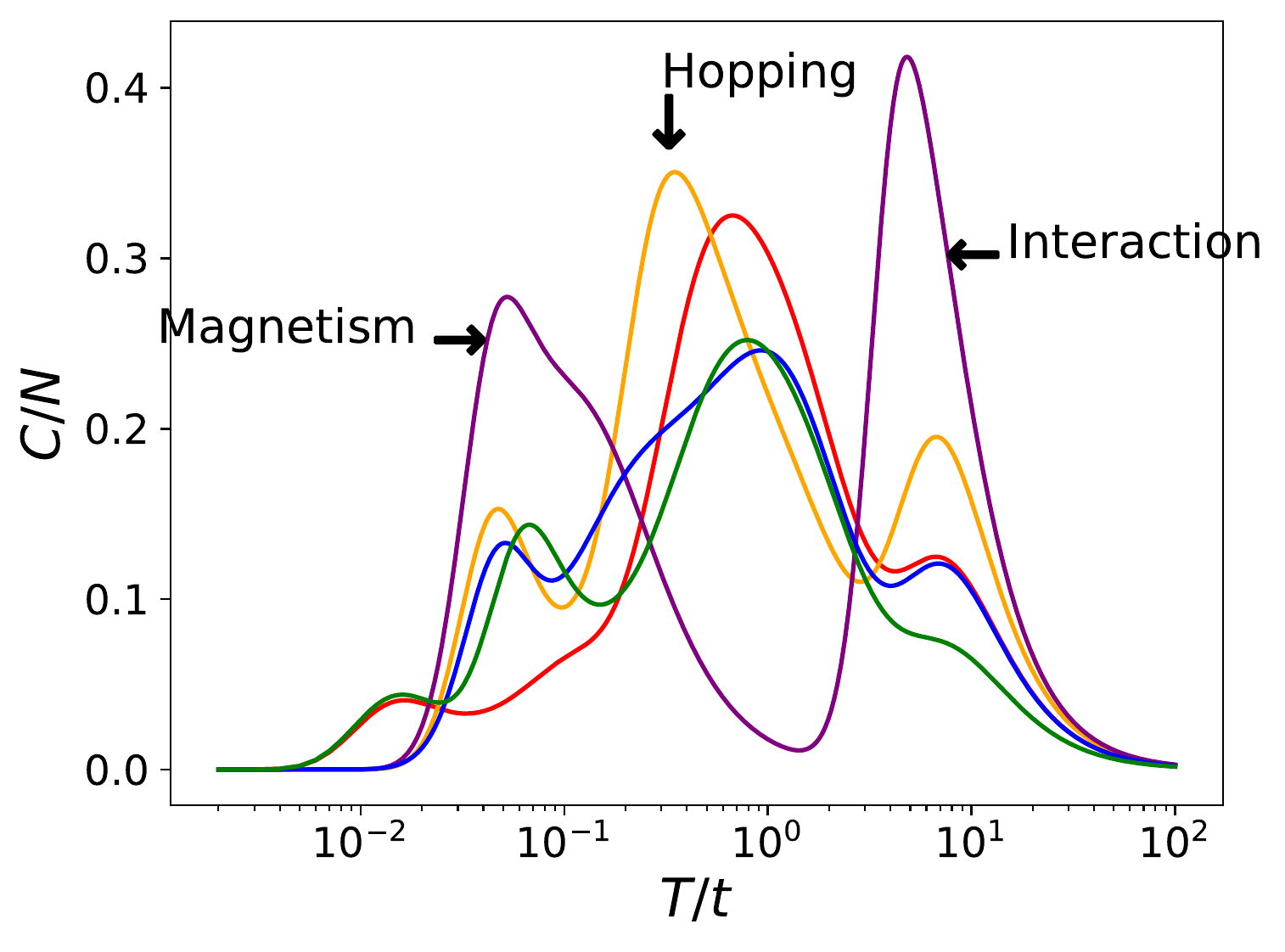}
\subfigimg[height=132pt,width=170pt]{%
\qquad\quad\textsf{\normalsize(c)}
}{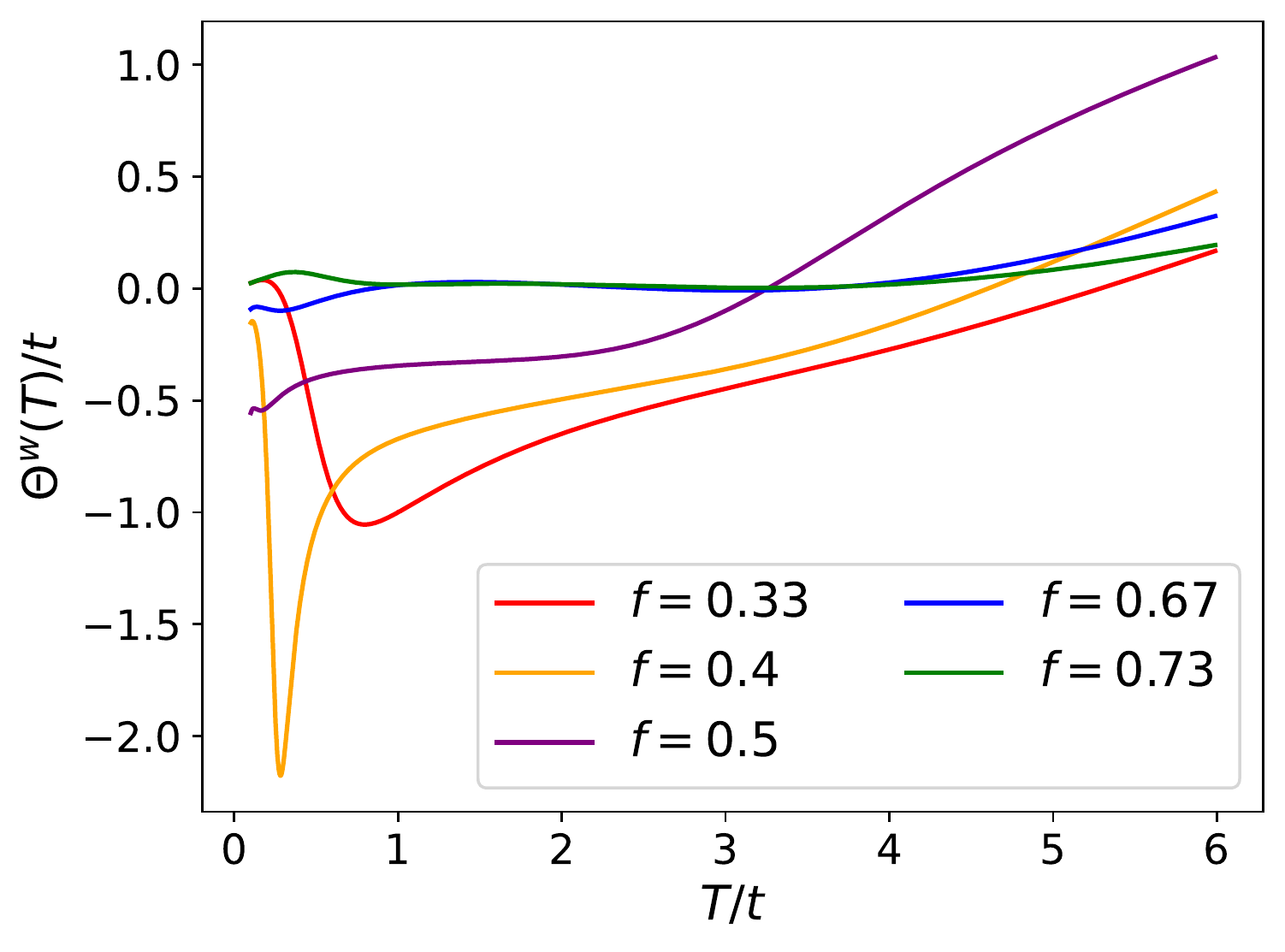}
\caption{(a) Curie-Weiss temperature ($\Theta$) as a function of density ($f$) for the triangular lattice Hubbard model with $U/t=20$. 
The simulations are compared to the experiments reported in Ref.~\cite{tang2020n}, denoted by Cornell.
(b) Specific heat ($C/N$) versus temperature of the T-15 cluster for representative fillings $f$ showing 
three regimes associated with magnetism, hopping and Hubbard interactions.
(c) Window-dependent Curie-Weiss temperature (as defined in the text) 
versus temperature for the T-15 cluster for representative fillings $f$. 
}
\label{fig:window_CW}
\end{figure*}

\para{}
Our work here is motivated by experiments on a transition metal dichalcogenide (TMD) $\text{WSe}_2/\text{WS}_2$ bilayer system~\cite{tang2020n, Wu_2018}, which realizes a  
triangular moire superlattice with a small hopping of approximately 10 Kelvin, with tunable density of holes and whose intermediate temperature scale 
has been readily accessed. We henceforth refer to Ref.~\cite{tang2020n} as the ``Cornell experiment" (CE). Rather curiously, and somewhat unexpectedly, CE 
reported an \textit{increase} in the absolute value of the Curie-Weiss (CW) temperature on reducing the hole density from half filling. This may appear counterintuitive 
and defies the expectation that the effective magnetic interactions must decrease (and hence the CW temperature must decrease) 
with lowering the particle density. CE also suggested the existence of a ferromagnet (FM) 
based on the positivity of the CW temperature for a range of densities above half filling. 
While the Nagaoka theorem~\cite{Nagaoka_1966,Thouless_1965}, admits such a possibility at infinite $U$ 
for the square lattice, FM has not been observed at finite $U$.
The frustrated/non-bipartite nature of the triangular lattice has been argued to potentially destabilize the tendency for local antiferromagnetism (AFM) and admit a FM ground state especially at large $U/t$~\cite{Hartmann1995,Hanisch_Hartmann,Krishnamurthy_triangular,Wang_triangular} and in a finite magnetic field~\cite{Davydova_Fu}. Nagaoka ferromagnetism has been found to be unstable in the $U=\infty$ limit on the hole doped side~\cite{Shastry_Anderson}, it is thus useful to clarify the theoretical situation for finite but large $U/t$.  

While the details of a quantitatively accurate effective Hamiltonian of the TMD bilayer system 
still remain to be completely fleshed out, we study the simplest model believed to be broadly consistent with experiments--%
the Hubbard model~\cite{Hubbard_1963} on the triangular lattice, 
\begin{align}
	H &= -t \sum_{\langle i, j \rangle, \sigma} \Big( c_{i,\sigma}^{\dagger} c_{j,\sigma} + c_{j,\sigma}^{\dagger} c_{i,\sigma} \Big) 
        + U \sum_{i} n_{i,\uparrow} n_{i,\downarrow}
\end{align}
where $c_{i,\sigma}^{\dagger} (c_{i,\sigma})$ refer to the usual creation (annihilation) operators with spin $\sigma$ at site $i$, and $n_{i,\sigma}$ refer to number operators.
$\langle i,j \rangle$ refers to nearest neighbor pairs of sites. We will denote the filling by $f\in [0,1]$. For example, $f=1/2$ refers to half filling, i.e., one hole per triangular site. 
We focus on the reported value of $U/t=20$~\cite{tang2020n} and intermediate temperatures $t \lesssim T < U$. 

A key result of this work is that our numerical simulations recover the reported trends of the CW temperature $\Theta$ as a function of particle density. 
 We also find the increase in $|\Theta|$ on reducing the particle density away from half-filling within the framework of the $t$-$J$ model, the low-energy limit of the Hubbard model. 
However, this trend 
is the \textit{opposite} of the one predicted by the high-$T$ expansion;  
we probe this further and explain the origin of this apparent conflict. 
Finally, we monitor the spin structure factor in our numerical calculations which explain how FM and AFM correlations develop on lowering the temperature. It also reveals parallels with the underlying zero temperature phase transition, which we explore with the ground state density matrix renormalization group (DMRG)~\cite{DMRG_White}.

\section{Finite temperature Lanczos results and comparison with experiments}
\label{sec:ftlm}
The ground state phase diagram of the triangular lattice Hubbard model has been extensively investigated, 
partly due to its relevance to organic-charge transfer salts such as (BEDT-TTF)$_2$X~\cite{Powell_2006},
using a variety of numerical methods including exact diagonalization (ED),
dynamical mean field theory (DMFT)~\cite{Li_PRB, Merino} and DMRG~\cite{Yunoki_triangularHubbard,Venderley_Kim,Zaletel_chiral, Weichselbaum_triangular}. 
At $f=1/2$ and for large $U/t$, 120 degree spiral order is stabilized; at low $U/t$, a metallic phase exists and at intermediate $U/t \approx 8$ a gapless~\cite{Yunoki_triangularHubbard} and possible chiral spin liquid~\cite{Zaletel_chiral} has been reported.
Less is definitively known for the case of doping away from half filling:
At low filling ($f=0.2-0.3$), a stripe AFM ~\cite{Li_PRB} and at higher filling ($f \sim 0.75$, and at large $U/t$) a FM is stabilized~\cite{Merino,Li_PRB}. 

\para{}
We explore the intermediate temperature regime~\cite{Wietek_PRX} with ED and the finite temperature Lanczos method 
(FTLM)~\cite{Jaklic_Prelovsek,Prelovsek_Bonca} on triangular lattices with $N = $ 9, 12 and 15 sites which we refer to as ``T-$N$" clusters 
(see Appendix~\ref{sec:clusters} for cluster shapes).
We typically use $M=150$ Krylov vectors and $R=100-1500$ seeds (per sector).
We compute the susceptibility $\chi$ (per site) at temperature $T$ within the framework of fluctuation dissipation, i.e., using 
\begin{equation}
	\chi = \frac{\langle S_z^2 \rangle_{\mathrm{th}} - {\langle S_z\rangle}_{\mathrm{th}}^2} {T N}
\end{equation}
where $\langle \cdots \rangle_{\mathrm{th}}$ represents the thermal average. 
Note that $\langle S_z \rangle_{\mathrm{th}} = 0$ since the Hamiltonian is time-reversal symmetric. 
To extract the CW temperature we use the mean field result, 
\begin{equation}
	\chi = \frac{C}{T - \Theta}
\end{equation}
where $C$ is the Curie constant (which for a purely magnetic model equals $\frac{1}{3} S(S+1)$ where $S$ is the spin of a single magnetic moment) 
and $\Theta$ is the CW temperature, $\Theta > 0$ corresponds to effective FM and $\Theta < 0$ AFM interactions.
In order to carry out a one-to-one comparison with CE,
we choose the same range of temperatures for fitting ($0.8 t \le T \le 5.5 t$).
The fitting range is important since the CW temperature is sensitive to the temperature window used, 
an issue we will elaborate on. 

\begin{figure*}
\subfigimg[height=125pt]{\qquad\qquad\;\;\textsf{\normalsize(a)}}
{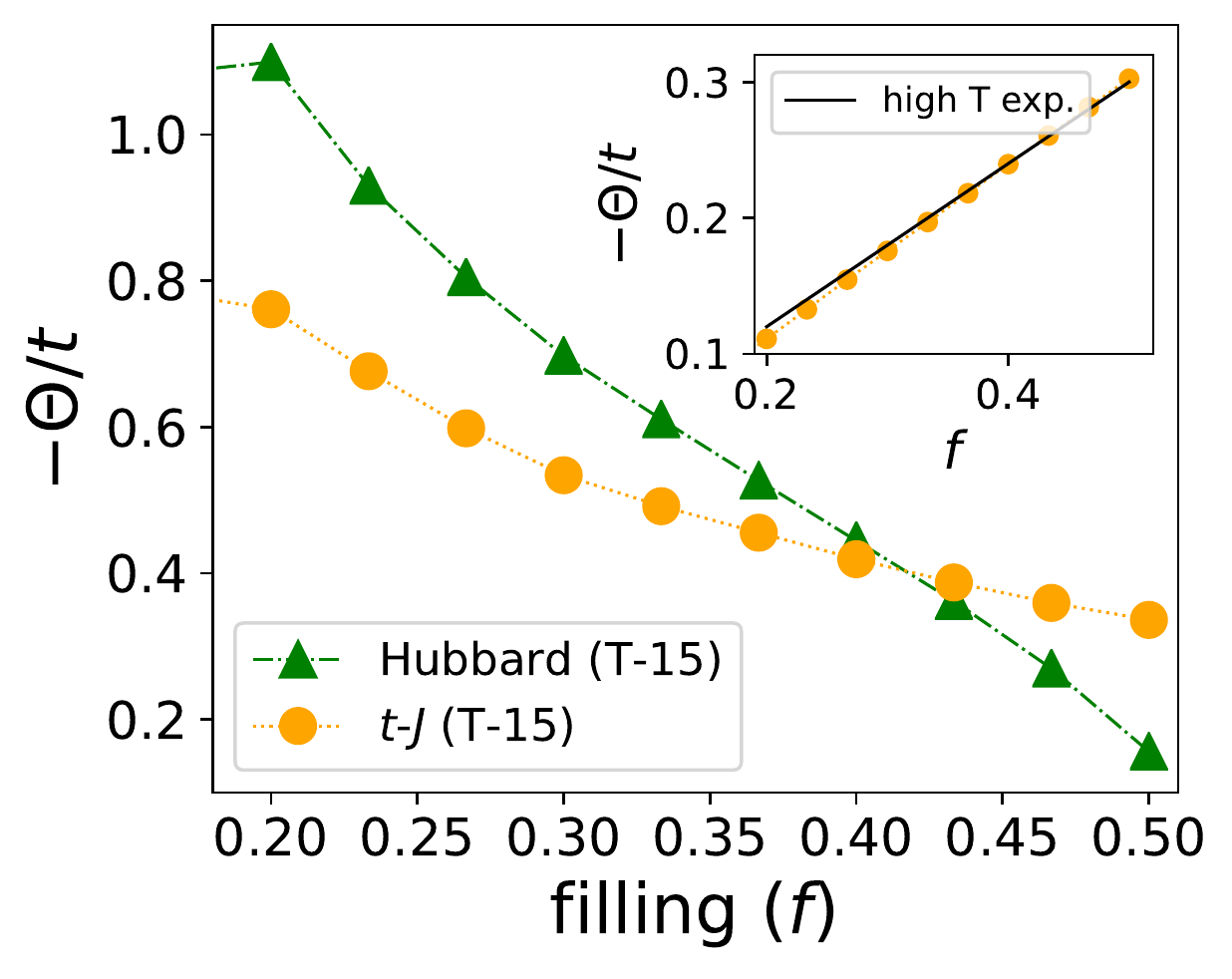}
\subfigimg[height=125pt]{\qquad\qquad\qquad\qquad\;\;\textsf{\normalsize(b)}}
{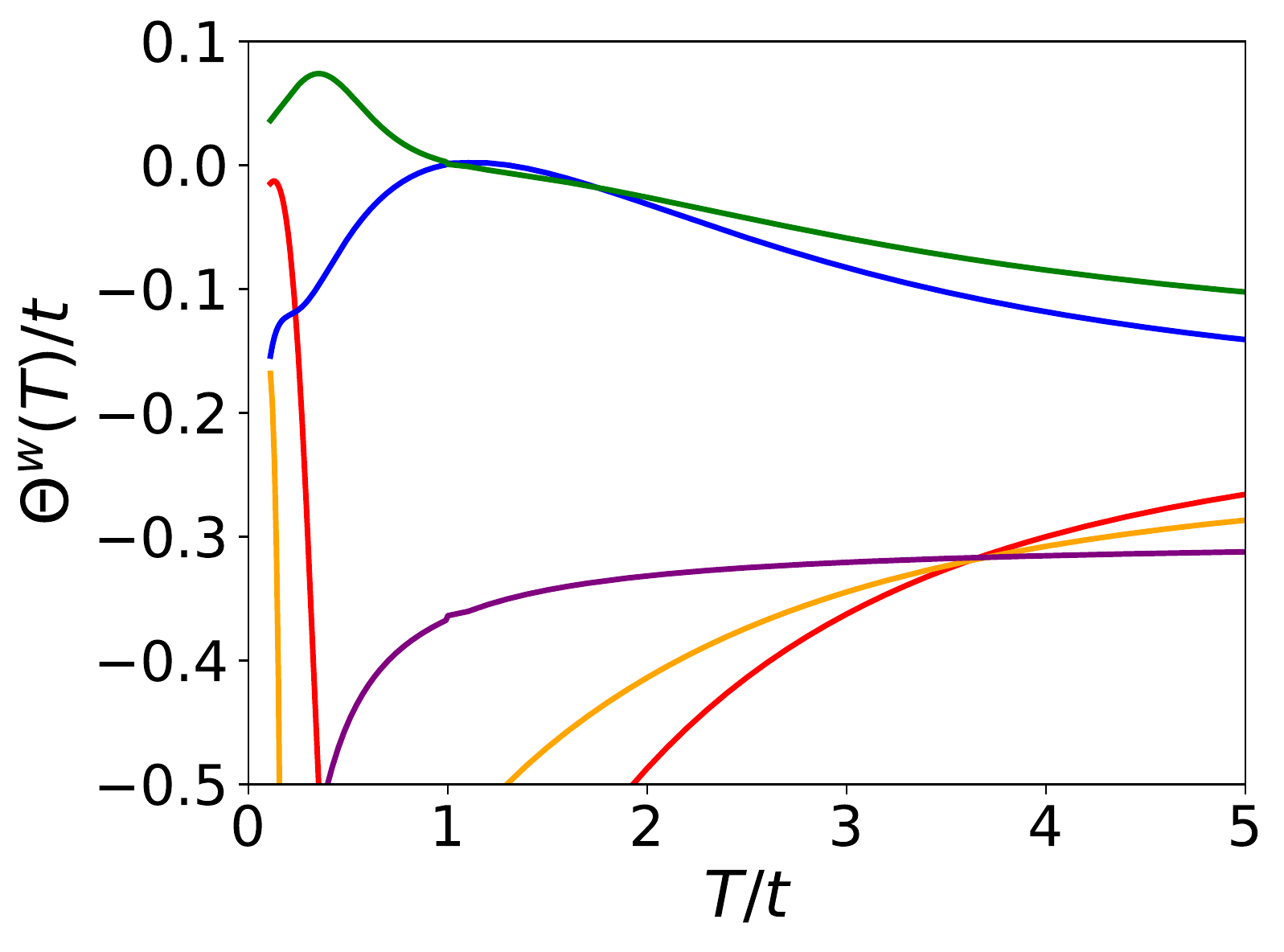}
\subfigimg[height=125pt]{\qquad\;\;\textsf{\normalsize(c)}}
{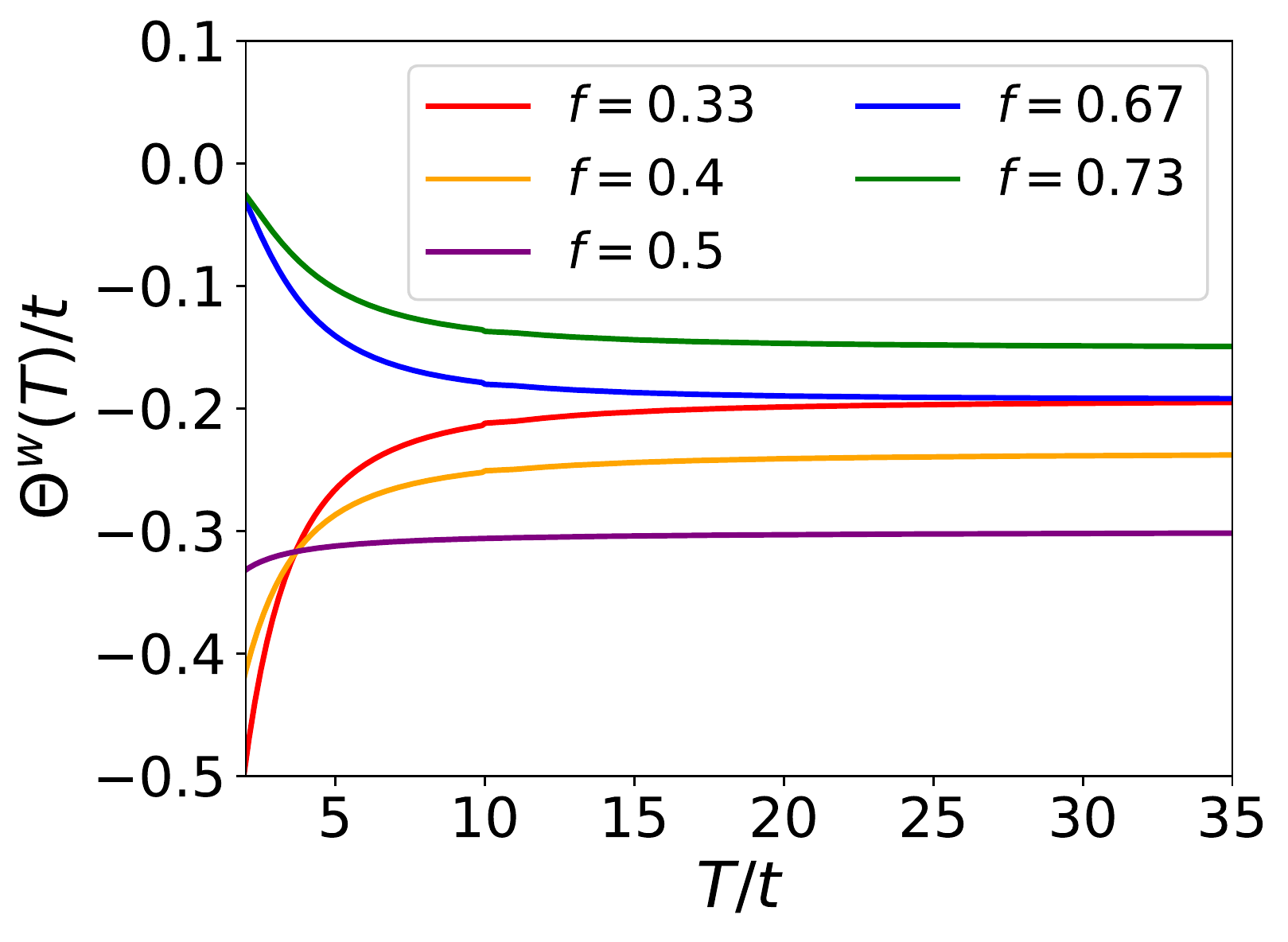}
\caption{
(a) Curie-Weiss temperature ($\Theta$) for the T-15 cluster as a function of $f$ for both the Hubbard and $t-J$ models using $U/t=20$ 
and $J/t=0.20$ respectively. The susceptibility was fitted in the temperature range $0.8t \le T \le 5.5t$. The inset shows $\Theta$ as a function of filling 
for the $t-J$ model, obtained by fitting the susceptibility data in the temperature range $20t \le T \le 30t$ and compared with the high-T series expansion result (to the lowest two orders).
Panels (b) and (c) show the window-dependent Curie-Weiss temperature ($\Theta^{\mathrm{w}}(T)$) of the triangular $t-J$ model, with the same parameters as in (a), in (b) the low to intermediate temperature and (c) the intermediate to high temperature regimes.}  
\label{fig:Hubbard_tJ}
\end{figure*}

\para{}
Results of our fits for two representative doping densities are presented in Fig.~\ref{fig:expt_theory} (other representative fits 
are shown in Appendix~\ref{sec:CWfits}). 
We find that the linear approximation for $1/\chi$ is indeed 
rather remarkable, at least visually, given that the experimental temperature ranges are not in the truly ``high temperature'' limit. 
We determine $\Theta$ from the intercept on the horizontal (temperature) axis, which is 
compared with CE in Fig.~\ref{fig:window_CW}(a). 
While there are variations in the CW estimates as a function of system size, 
the general trends and magnitude of the effect are captured well, given the inherent experimental uncertainties. 
For the purpose of comparison, we have also considered the case of the square lattice Hubbard model in Appendix~\ref{sec:square}. 

\para{}
The susceptibility is characterized by the CW form only for temperatures much higher than the effective magnetic interaction $J$. The situation is more complex for the Hubbard model - 
there are three energy scales: $J \sim t^2/U$ which is related to magnetism, effective/renormalized 
hopping/kinetic energy (bandwidth), and Hubbard interactions. 
The specific heat, shown in Fig.~\ref{fig:window_CW}(b) for the T-15 cluster, 
reveals these scales. For example, the 
effective hopping is quenched at $f=1/2$, but it does show up as a intermediate temperature bump that (typically) moves to higher temperature on lowering the filling, 
by $f=1/3$ this bump occurs at $T \sim t$.
A third bump at higher temperature corresponds to the scale at which double occupancy becomes important 
, we observe that this scale also increases on either side of half filling.
This filling dependence is expected since (for $f<1/2$) the lower the density, the easier it is to avoid the double occupancy cost at increasing temperatures. This temperature scale is less than $U=20 t$, in fact, double occupancy fluctuations are considerable at the highest temperature probed in CE ($T \approx 6t$) and used in the CW fitting, which we comment on further in the next section. Due to the presence of these three distinct energy scales in the Hubbard model, the low energy physics ($T\lesssim t^2/U$) is dominated by magnetic interactions and we observe competition among different magnetic orders, the intermediate scale ($T\sim t$) is dominated by phenomena associated with  kinetic frustration~\cite{Haerter_Shastry_2006}, and high temperature ($T\gtrsim U$) by charge fluctuations. 

\para{}
Since the CW theory is based on the properties of a magnetic model in its high temperature regime, the $\Theta$ extracted corresponds to the best mean-field fit 
which depends sensitively on the window of temperature used in its determination. To make these notions precise, we define the ``window-dependent'' 
CW temperature, $\Theta^{\mathrm{w}} (T)$, obtained by extrapolating the inverse susceptibility at a given temperature all the way to zero,
\begin{align}
	\Theta^{\mathrm{w}} (T) \equiv T - \chi^{-1} \Big( \frac{d \chi^{-1}}{d T} \Big)^{-1}.
\end{align}
Figure~\ref{fig:window_CW}(c) shows the variation of $\Theta^{\mathrm{w}}$ with temperature for $T \le 6t$. 
As expected, there is a big variation at low temperature, however, 
even for $T \gtrsim t$ we find that $\Theta^{\mathrm{w}}$ is not flat, as can be prominently seen for $f=0.33,0.4$ and $0.5$. 
It reveals that $\chi^{-1}$ is not perfectly linear with temperature, and thus 
the reported $\Theta$ reflects an average value in the specified temperature window. 

\section{Insights from the $t-J$ model}
\label{sec:tJ}
For $T \gg \Theta$, the CW theory can be thought of as a series 
expansion for $\chi$ in powers of $1/T$ (by Taylor expanding $1/(T-\Theta)$), which can be compared with 
high temperature series expansions. The term proportionate to $1/T$ gives the paramagnetic susceptibility, 
while the next order term gives $\Theta$. For $ T < U$ it is convenient to work within
the framework of the $t$-$J$ model, the low-energy limit of
the Hubbard model for large $U/t$
~\cite{Chao_Spalek,FazekasBook}. 
Its Hamiltonian is,
\begin{align}
	H &= -t \!\! \sum_{\langle i, j \rangle, \sigma} \!\! P \Big( c_{i,\sigma}^{\dagger} c_{j,\sigma} + c_{j,\sigma}^{\dagger} c_{i,\sigma} \Big) P
	+ J \sum_{\langle i,j \rangle} \left( \bfS_{i} \cdot \bfS_{j} - \frac{1}{4} n_{i} n_{j} \right).
\end{align}
The first term is a ``restricted hopping,'' i.e., one which never permits two holes to be on the same site.
In other words, $P$ projects out states with one or more doubly occupied sites for $f \le 1/2$, and empty sites in the case of $f \ge 1/2$. 
The second term corresponds to magnetic Heisenberg interactions and density-density interactions arising from degenerate perturbations of the restricted manifold. $|\Theta|$ from the high-$T$ expansion of the $t$-$J$ model is $\frac{z}{4}J f$~\cite{Singh_Glenister} where $z$ is the coordination number 
($z=4$ and $z=6$ for the square and triangular lattice respectively),
i.e., it must \emph{decrease} as one decreases $f$ from 1/2~\cite{Zhang_Yuan_Fu}, a result which is at complete odds with CE.

So how should one reconcile these apparently contradictory findings? 
The resolution to this puzzle lies in the fact that the high temperature and intermediate temperature regimes of the $t-J$ model are not the same, even qualitatively. 
(The high temperature regime is on a scale of $U$ or larger and is thus not of direct relevance to what is measured in CE.) We check our assertion 
by exploring the CW temperature of the $t-J$ model in the experimentally relevant intermediate temperature range; 
Fig.~\ref{fig:Hubbard_tJ}(a) shows that the $t-J$ model captures the same trends as the Hubbard model and CE. (For more comparisons between the two models see Appendix~\ref{sec:apptJ}.) 
In sharp contrast, and in perfect quantitative agreement with the high-T expansion result, 
the inset of Fig.~\ref{fig:Hubbard_tJ}(a) shows the decrease in CW temperature 
of the $t-J$ model on reducing $f$ from $1/2$, captured by our FTLM results by using the fitting range $20t \leq T \leq 30 t$.

This calls for a careful look at the temperature window-dependence of the CW temperature for the $t-J$ model. 
We plot $\Theta^{\textrm{w}}$ for the $t-J$ model 
in the low to intermediate temperature regime and in the intermediate to high temperature regime in Fig.~\ref{fig:Hubbard_tJ}(b) and (c) respectively. 
We find enhancement of $\Theta^\textrm{w}$ on decreasing $f$ for $f<1/2$ in the intermediate temperature regime. This trend clearly changes on going to the high temperature regime - 
this crossover occurs at $T \approx 3.7t$. For $f \approx 0.73$ we observe a positive $\Theta^\textrm{w}$ at low temperature, which at intermediate temperature crosses over to a negative value. 
Also note that at extremely high temperature, $\Theta^\textrm{w}$ is particle- hole symmetric about half filling, see for example $f=0.33$ and $f=0.67$ in Fig.~\ref{fig:Hubbard_tJ}(c), even though the underlying triangular lattice $t-J$ Hamiltonian does not have that symmetry. This is because at high temperatures, only on-site and nearest neighbor correlations dominate and thus information about the underlying lattice structure (including loops) is greatly suppressed.

As one cools the system, other correlations begin to contribute to $\chi$ and the nature of the lattice (e.g. frustrated or not) becomes important. The significance of frustration for doped magnets was realized in pioneering work by Haerter and Shastry~\cite{haerter2005prl} who studied thermodynamics of the $t-J$ model on the triangular lattice in the context of sodium cobaltate, 
which resulted in the theory of kinetic frustration~\cite{haerter2005prl, Sposetti_2014}. This theory can be summarized as follows. Consider $U \rightarrow \infty$ in the 
$t-J$ model which corresponds to $J=0$. For $f=1/2$ all magnetic orders are exactly degenerate, since magnetism is completely suppressed. When a single particle is removed or 
added on a square lattice, the kinetic term (proportionate to $t$) favors the hole or doublon to move in a FM background. However, for the triangular lattice, which lacks particle-hole symmetry 
the result is very different - removal of a particle favors (120 degree) AFM and addition favors FM. Thus even in the absence of any magnetic interactions, 
the kinetic energy prefers an AFM state, at least at low doping.

\begin{figure*}
\subfigimg[height=175pt]
{\textsf{\normalsize(a)} \qquad\qquad\qquad\qquad\qquad\qquad\qquad\qquad\qquad\qquad\qquad\qquad\qquad\qquad\qquad \;\;\;\;\;\;\; \textsf{\normalsize(b)}}
{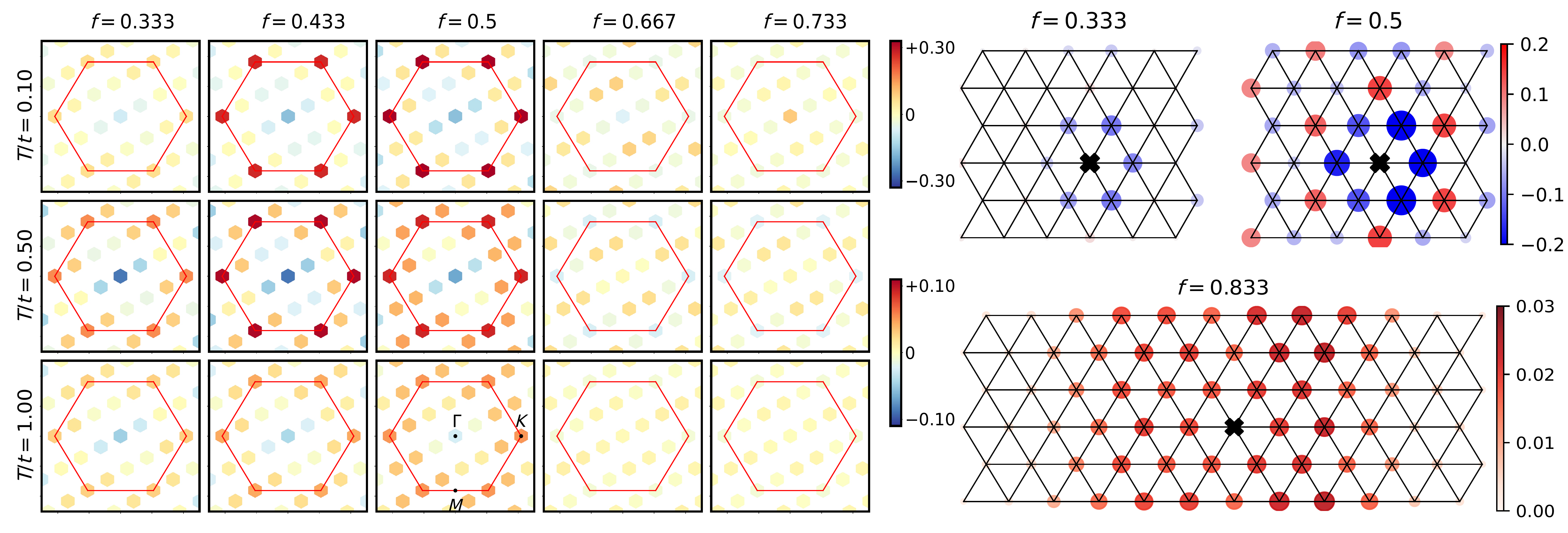}
\caption{
(a) Static structure factor with a high temperature subtraction $S^{zz}(\bfq,T)- S^{zz}(\bfq,5t)$, for the triangular Hubbard model ($U/t=20$) across various fillings $f$ for the T-15 cluster for $T/t=0.1,0.5,1$. $M=150$ Krylov vectors and $R=300$ seeds were used for the FTLM. 
The red hexagon in each panel marks the Brillouin zone boundary, and prominent momentum points (${\bf \Gamma, K, M}$) are indicated.
(b) DMRG ground state real space spin-spin correlations $\langle \textbf{S}_o \cdot \textbf{S}_l\rangle $ for every lattice site ($l$) with respect to a centrally chosen site ($o$) marked with black cross for $f=1/3,1/2$ on a length 6 and $f \approx 0.833$ on a length 12 XC-6  cylinder. The diameters of the circles are proportional 
to the amplitudes of the spin-spin correlation and the colors indicate the sign of the correlations. 
}
\label{fig:S_q}
\end{figure*}

These arguments strictly hold at $U \rightarrow \infty$ but 
should apply, with some modifications, to the case of large but finite $U$.
For finite $U$, i.e. non-zero $J$, the low temperature regime ($T < J$) is dominated by the competition between different magnetic orders. At $f=1/2$ and large $U/t$ the 120$^{\circ}$ antiferromagnetic state is selected, while close to $f \approx 0.75$, an itinerant Stoner ferromagnet is favored, see Fig.~\ref{fig:S_q}.  
Antiferromagnetic correlations, characterized by the strength of the weight at the ${\bf K}$ points in the Brillouin zone, are expected to weaken on lowering $f$ from $1/2$ - this assertion will be substantiated in the next section. 
However, at intermediate temperatures ($U>T>J$) and $f=1/2$ this competition between magnetic states is greatly suppressed, in this regime one can think of all the competing magnetic states as essentially degenerate with one another. The dominant scales in this temperature regime are then only $t$ and $U \gg t$ (which can be thought of as infinitely large) and it is in this regime that the Haerter-Shastry arguments should apply i.e. one should expect the kinetic energy to enhance AFM in this temperature range. This expectation is 
borne out in CE and our numerical data. 

\section{Finite temperature remnants of AFM to FM transition} 
\label{sec:ssf}
To further explore the enhancement of AFM correlations we study, with FTLM, the thermal momentum dependent spin structure factor (SSF) 
\begin{equation}
	S^{zz}({\bf q},T) \equiv \frac{1}{N} \sum_{i,j} e^{-i\bfq \cdot (\bfr_i - \bfr_j)} \langle S^{z}_i S^{z}_j \rangle_{\mathrm{th}}
\end{equation}
where $\bfr_i$ represents the physical coordinate of site $i$. Note that the SSF is equivalent to $T\chi(\bfq,T)$, $\chi(\bfq=(0,0),T)$ corresponds to the measured susceptibility. 
Though limited by obvious finite size effects, our calculations shed insights on various competing orders especially at small correlation lengths, i.e., higher temperature. We also address the small positive $\Theta$ that signals FM,
which is weak and possibly fragile as suggested by the FTLM calculations ($\Theta>0$ is captured only on the largest cluster we studied, 15 sites).
A previous DMFT (dual-fermion) study in the low-temperature limit ($T=0.1t$) has reported the presence of FM~\cite{Li_PRB} up to $U/t=10$, here we focus 
on $U/t=20$ and additionally explore the relationship between low and intermediate temperatures. 

\para{}
We subtract out the high temperature correlations $S^{zz}(\bfq, T=5t)$ and plot the difference in Fig.~\ref{fig:S_q} for the T-15 cluster for $U/t=20$ across different 
fillings at $T/t=0.1, 0.5$ and $1$. The importance of subtracting out the high temperature data is clarified in Appendix~\ref{sec:ssf_other}. 
The T-15 cluster retains the prominent momentum points (and their symmetry related partners): ${\bf K} = (\frac{4\pi}{3},0)$ 
(that captures the 120 degree N\'eel ordering) and ${\bf \Gamma}=(0,0)$ (that captures FM), but not ${\bf M} = (0, \frac{2\pi}{\sqrt{3}})$. 
(Additional results are shown in Appendix~\ref{sec:ssf_other} for the T-12 cluster which does contain the {$\bf M$} point.) 
For a given filling, the subtracted SSF monitors the tendency for formation of magnetic orders on cooling the high temperature state. Strictly speaking, there is no true long range order in two dimensions at \textit{any} finite temperature due to the Hohenberg-Mermin-Wagner theorem.\cite{mermin1966prl,hohenberg1967pr}. 

For $f=1/2$ we see development of weight at the ${\bf K}$ points on cooling. Importantly,
at low temperature ($T \lesssim J=0.2 t$, see for example, $T=0.1t$ in Fig.~\ref{fig:S_q}(a)), 
AFM correlations are weakened on doping (decreased redness at ${\bf K}$ points), as one may intuitively expect.
For $f \gtrsim 1/2$, the weight at the ${\bf K}$ points at low temperature is lost eventually migrating towards the ${\bf \Gamma}$ point signaling the onset of FM correlations. (At low temperature, the overall weak scale of FM relative to the AFM is apparent from the redness of the color at the ${\bf \Gamma}$ vs ${\bf K}$ points.) Prominently, at intermediate temperature ( $T=0.5 t$ and $T=t$) there is an enhancement of weight for $f=0.433 $ at the ${\bf \Gamma}$ point relative to $f=1/2$ (it gets bluer), qualitatively consistent with the increase in the CW temperature reported by CE. 
There is also a mild enhancement at the {\bf K} points for $T=0.5 t$. 

\para{}
For the $T=0$ case (where finite size effects are most prominent) we performed ground state DMRG on XC-6 cylinders of length 6 and 12 (36 and 72 sites respectively) retaining up to 16000 states. Figure~\ref{fig:S_q}(b) shows the results of the real space spin-spin correlation functions with respect to a centrally chosen site for $f=1/3$, $f=1/2$ and $f \approx 0.833$. The case of $f=1/3$ exhibits extremely short range AFM nearest neighbor correlations. 
For $f=1/2$ and $f \approx 0.833$, the qualitative conclusion from FTLM holds: The correlations are clearly AFM (longer-range) and FM respectively. 
The real space pattern of spin-spin correlations for $f=1/2$ closely resembles what was previously observed for the spin-1/2 Heisenberg model~\cite{Pal_Sharma_colorful}. 
The momentum dependent static structurefactor for the ground states at representative fillings has been shown in Appendix ~\ref{sec:appssf}.

Due to the closeness of the FM to the van Hove singularity at $f=0.75$, the appearance of FM at high fillings is expected to be due to a Stoner instability. We find a reduced magnetic moment, for example for $U/t=20$ and $f \approx 0.833$ we find the moment to be roughly half of what would be expected for a fully polarized FM at the same particle density. Due to the effectively low density of spin carrying particles (doublons do not contribute to the magnetic moment) in this regime, and the reduced moment from quantum fluctuations, the FM correlations are weak compared to the corresponding AFM counterparts. This is at the heart of the small $\Theta$ observed in CE. Note however that CE sees FM at possibly lower $f \sim 0.6-0.7$, but also reports a considerable errorbar in $f$ of $0.1$. This requires a further review of both the model and the experiments, in particular it would be valuable to precisely nail down the extent and location of FM in the triangular Hubbard model. We leave the resolution of this and related issues to future work.

\section{Conclusion}
\label{sec:conclusion}
In summary, we have studied the intermediate-temperature physics associated with the triangular lattice Hubbard model, 
and reproduced several aspects of the Cornell experiment on the moire superlattice formed by WS$_2$ and WSe$_2$~\cite{tang2020n}.
In general, however, we expect the need for more refined models of moire materials~\cite{tang2022}.
We emphasize that increase in $|\Theta|$ on lowering filling \textit{does not necessarily} imply the strengthening of magnetic correlations in the ground state. We interpret the experimental and numerical results in the intermediate temperature regime within the framework of kinetic frustration which has been shown to enhance antiferromagnetism on doping~\cite{Haerter_Shastry_2006, Sposetti_2014}. We emphasize that there are prominent differences between low, intermediate and high temperature behaviors, this was demonstrated in the context of the $t-J$ model. 
We showed that the high temperature limit of the $t-J$ model gives a trend of CW temperatures with particle density that is the opposite of the trend observed in the intermediate temperature regime. We also studied the momentum-dependent structure factor as a function of temperature to clarify the trends in the susceptibility (associated with the ${\bf \Gamma}$ point) and the 120 degree magnetic ordering (associated with the ${\bf K}$ points). 

Using a combination of FTLM and DMRG calculations, we explored the possibility of FM in a regime of fillings where $\Theta>0$ was observed in the Cornell experiment. We found evidence in favor of a FM ground state that leaves its signature at finite temperature, consistent with previous work with complementary techniques~\cite{Li_PRB,Merino}. (There appears to be some disagreement on the precise extent and location of the FM in existing phase diagrams~\cite{Li_PRB, Merino}, it would be desirable for future work to clarify this issue.) The weak spin-spin correlations seen in our calculations offer an explanation of the smallness of the observed CW temperature. Similar observations have also been noted in the context of a recent cold atom experiment~\cite{Xu_Greiner_2022} which realizes a doped triangular Hubbard model. 

More generally, our work highlights the usefulness of comparing the results of many-body calculations with those of analog simulators, in this case a moire superlattice system formed by WSe$_2$ and WS$_2$. These simulators give access to a part of phase space, here intermediate temperatures, that may not be accessible to conventional materials thereby revealing new physics beyond the usual low energy, low temperature regime.

\para{Note added} At the time of submission of the first version of this paper, we became aware of a parallel preprint~\cite{Morera_2022} which has addressed similar questions.

\section*{Acknowledgments}
We thank K. Yang, V. Dobrosavljevic, K. F. Mak, S. Shastry, I. Morera, E. Demler, C. Chung, and M. Davydova for insightful discussions and J. Shan for a condensed matter 
seminar at NHMFL in 2020 that inspired some of the questions posed here. We thank A. Bhardwaj for discussions on high temperature series expansions. 
P.S. and H.J.C. were supported by NSF CAREER grant DMR-2046570.  O.V. was supported by NSF Grant No. DMR-1916958. K.L. and H.J.C. thank Florida State University and the National High Magnetic Field Laboratory for support. 
The National High Magnetic Field Laboratory is supported by the National Science Foundation through NSF/DMR-1644779 and the state of Florida. 
The numerical computations were carried out on resources provided by the Research Computing Center (RCC) and 
the Planck and Landau clusters at Florida State University. We thank A. Volya and U.S. Department of Energy, Award Number DE-SC-0009883 for additional computational resources.
The code used for finite temperature Lanczos calculations is available at Ref.~\cite{HubbardFTLM}.
The DMRG calculations were performed with the ITensor software~\cite{itensor}. 

K.L. and P.S. contributed equally to this work.

\appendix
\section{Finite Clusters for the Exact Diagonalization Calculations}
\label{sec:clusters}
\begin{figure}
\includegraphics[width=0.35\linewidth]{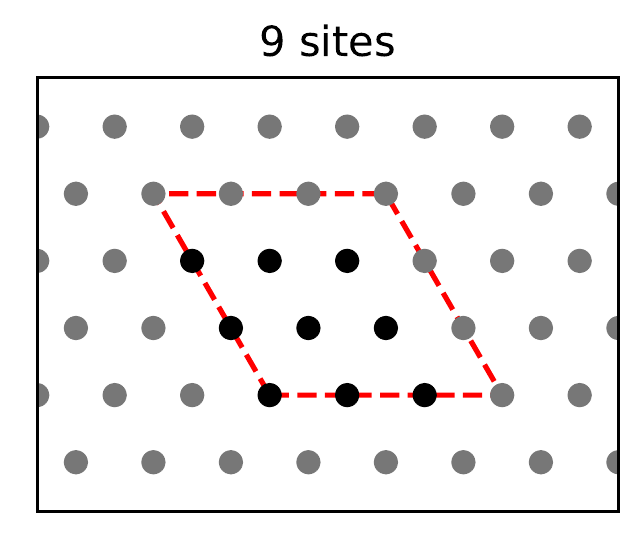}   
\includegraphics[width=0.22\linewidth]{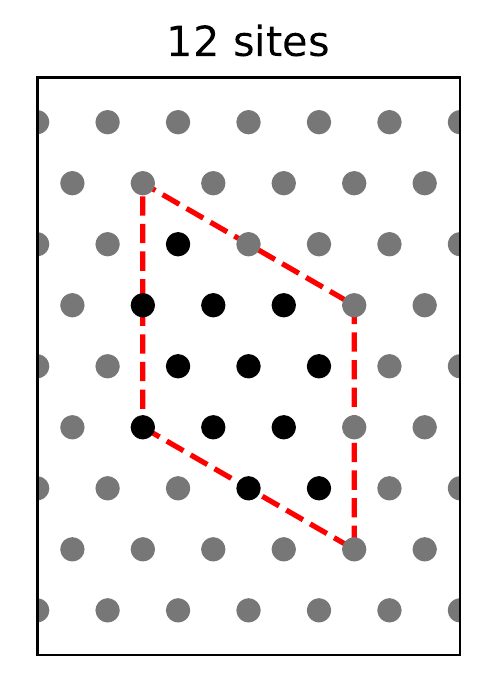}  
\includegraphics[width=0.35\linewidth]{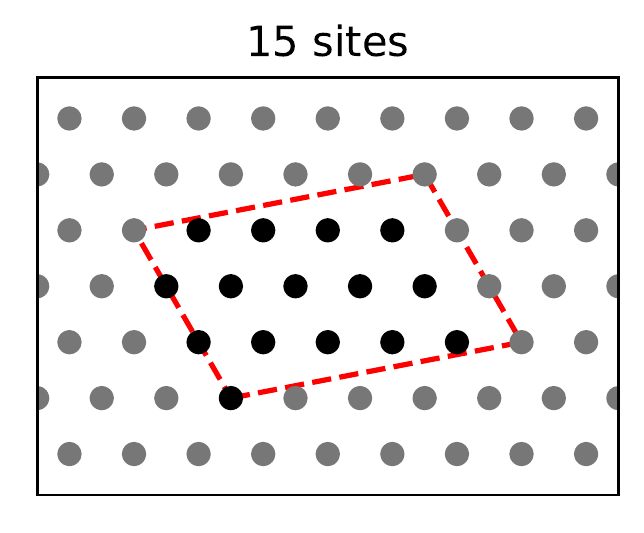}
\includegraphics[width=0.32\linewidth]{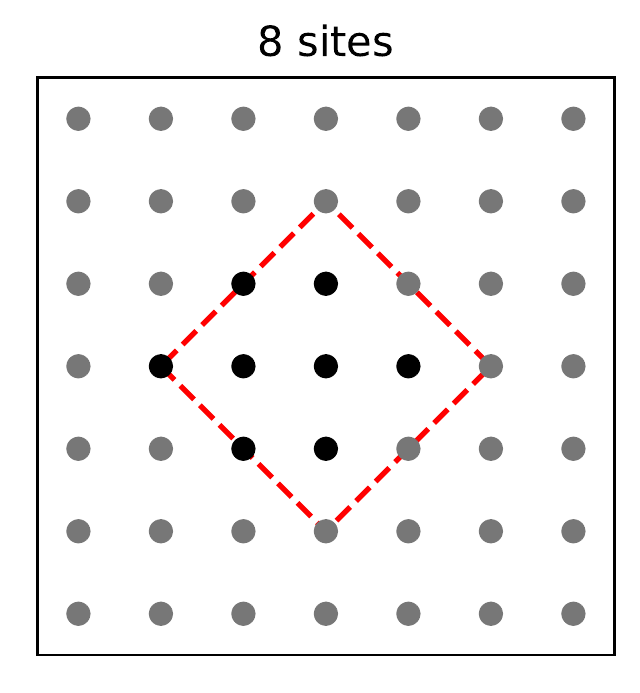} 
\includegraphics[width=0.32\linewidth]{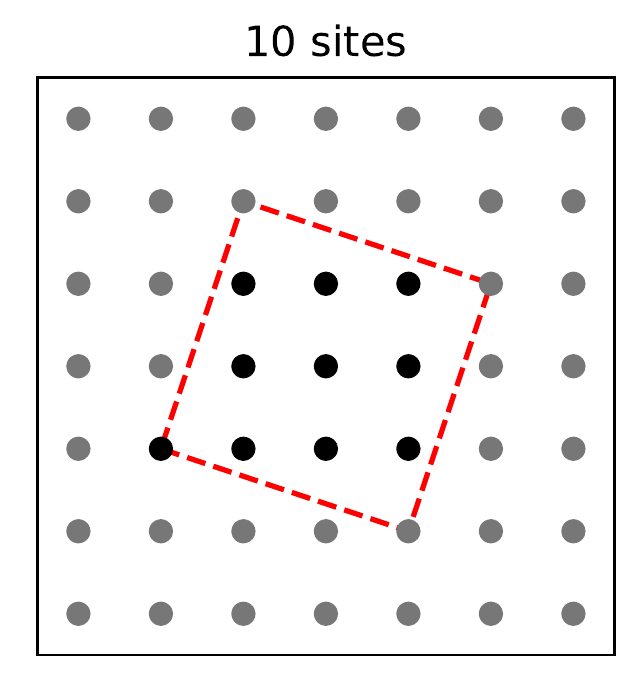}
\includegraphics[width=0.32\linewidth]{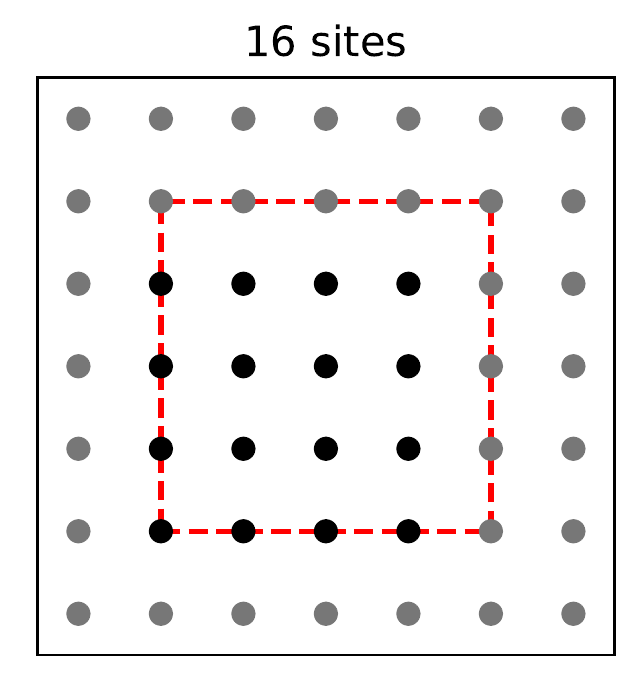}
\caption{\label{fig:lattices}
Finite triangular and square clusters treated with ED or FTLM in the main text and appendices.}
\end{figure}

In the main text and appendices we have presented results of ED and FTLM calculations. The finite clusters are shown in Fig.~\ref{fig:lattices}, 
they are frequently referred to as ``S-" (square) or ``T-" (triangular) followed by the number of sites. For example, S-10 is the 10 site square cluster, and T-15 is the 15 site triangular cluster.

\begin{figure}
\includegraphics[width=\linewidth]{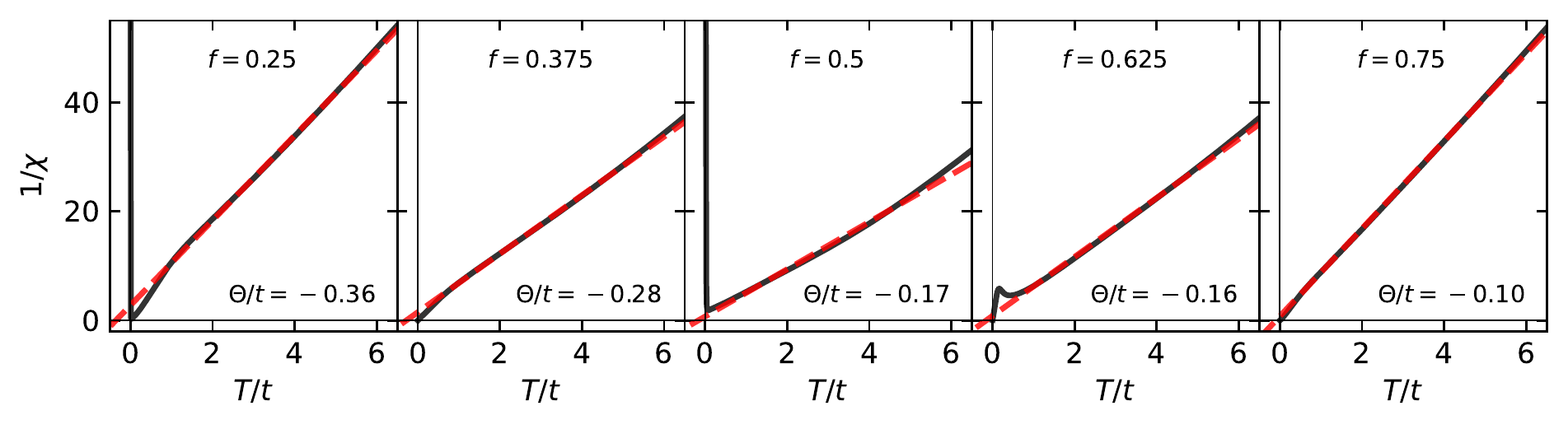}
\caption{Curie-Weiss fits to the inverse susceptibility (using a temperature range similar to that studied in the Cornell experiment~\cite{tang2020n}) for the T-12 cluster for various representative fillings.}
\label{fig:CWfits}
\end{figure}

Periodic boundary conditions were chosen for all simulations in this work. 
The momentum points (${\bf q}$) that are allowed by translational symmetry are determined by setting $e^{i {\bf q} \cdot {\bf R}}=1$ where ${\bf R}$ 
is the lattice vector associated with the periodicity of the cluster.
For example, for the T-12 cluster we have (in units of the lattice constant)
${\bf R_1} = 2 \sqrt{3} \hat{y}$ and 
${\bf R_2} = 3 \hat{x} - \sqrt{3} \hat{y}$,
which gives, 
\begin{subequations}
\begin{eqnarray}
    q_x &=& \frac{ (2 n + m) \pi}{3} \\
    q_y &=& \frac{ m \pi} {\sqrt{3}}
\end{eqnarray}
\end{subequations}
where $m,n$ are integers. For the T-15 cluster ${\bf R_1} = -\frac{3}{2} \hat{x} + \frac{3\sqrt{3}}{2} \hat{y}$ and ${\bf R_2} = \frac{9}{2} \hat{x} + \frac{\sqrt{3}}{2} \hat{y}$, gives,
\begin{subequations}
\begin{eqnarray}
    q_x &=& \frac{2 \pi}{15} (3 n - m) \\
    q_y &=& \frac{2\pi}{5\sqrt{3}} (3m+n)
\end{eqnarray}
\end{subequations}
It follows that the T-12 cluster has both ${\bf K} = (\frac{4 \pi}{3}, 0)$ and ${\bf M} = (0, \frac{2 \pi}{\sqrt{3}}) $ (and symmetry related) points whereas the T-15 cluster has only the ${\bf K}$ points but not the ${\bf M}$ points in its first Brillouin zone. Both clusters have the ${\bf \Gamma} = (0,0)$ point.

\section{Curie-Weiss fits for the triangular lattice}
\label{sec:CWfits}
In Fig.~\ref{fig:CWfits} we show representative Curie-Weiss (CW) fits to the inverse magnetic susceptibility (per site) for the T-12 cluster.

\begin{figure}
\includegraphics[width=0.8\linewidth]{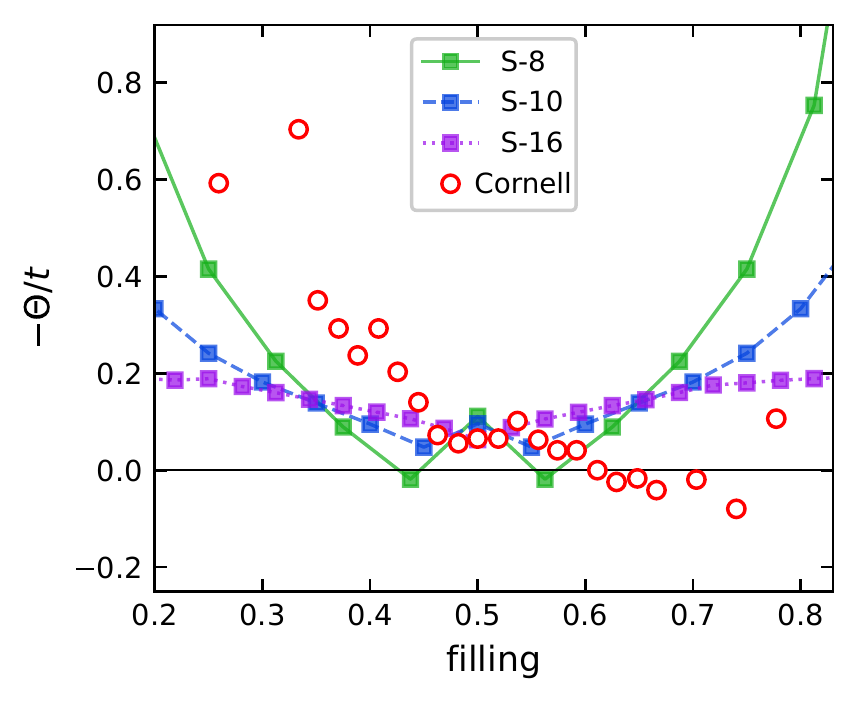}%
\caption{
Curie-Weiss temperature ($\Theta$) versus filling for the square lattice Hubbard model with $U/t=20$ (for three finite sizes) as compared to the Cornell experiment~\cite{tang2020n}.
}
\label{fig:CWsquare}
\end{figure}

\section{Curie-Weiss temperature for the Square Lattice Hubbard Model}
\label{sec:square}
In Section~\ref{sec:ftlm} we discussed the CW temperature $\Theta$ for the triangular lattice Hubbard model with nearest-neighbor hoppings as a function of (hole) filling.
Interestingly, this simple model admits a positive CW temperature, corresponding to FM, 
consistent with findings of CE~\cite{tang2020n}.
To provide a comparative check, we carried out numerical calculations for the square lattice case.

Figure~\ref{fig:CWsquare} shows results for the CW temperature for S-8, S-10 and S-16 with $U/t=20$. 
The CW fits were performed in a temperature range $0.8 < T/t < 5.5$, similar to the range chosen in CE. 
We find that $\Theta<0$ for all fillings, corresponding to effective antiferromagnetic (AFM) interactions. The exception is $8$ sites, where $\Theta>0$ for two fillings (related by particle-hole symmetry of the square lattice Hubbard model);
for larger system sizes, this tendency goes away. 
Additionally, for our largest size (S-$16$) the magnitude of the increase of the CW temperature on going from half filling towards lower filling 
is smaller than that observed in CE.

\begin{figure}
\subfigimg[height=106pt]{\qquad\qquad\;\;\textsf{\normalsize(a)}}
{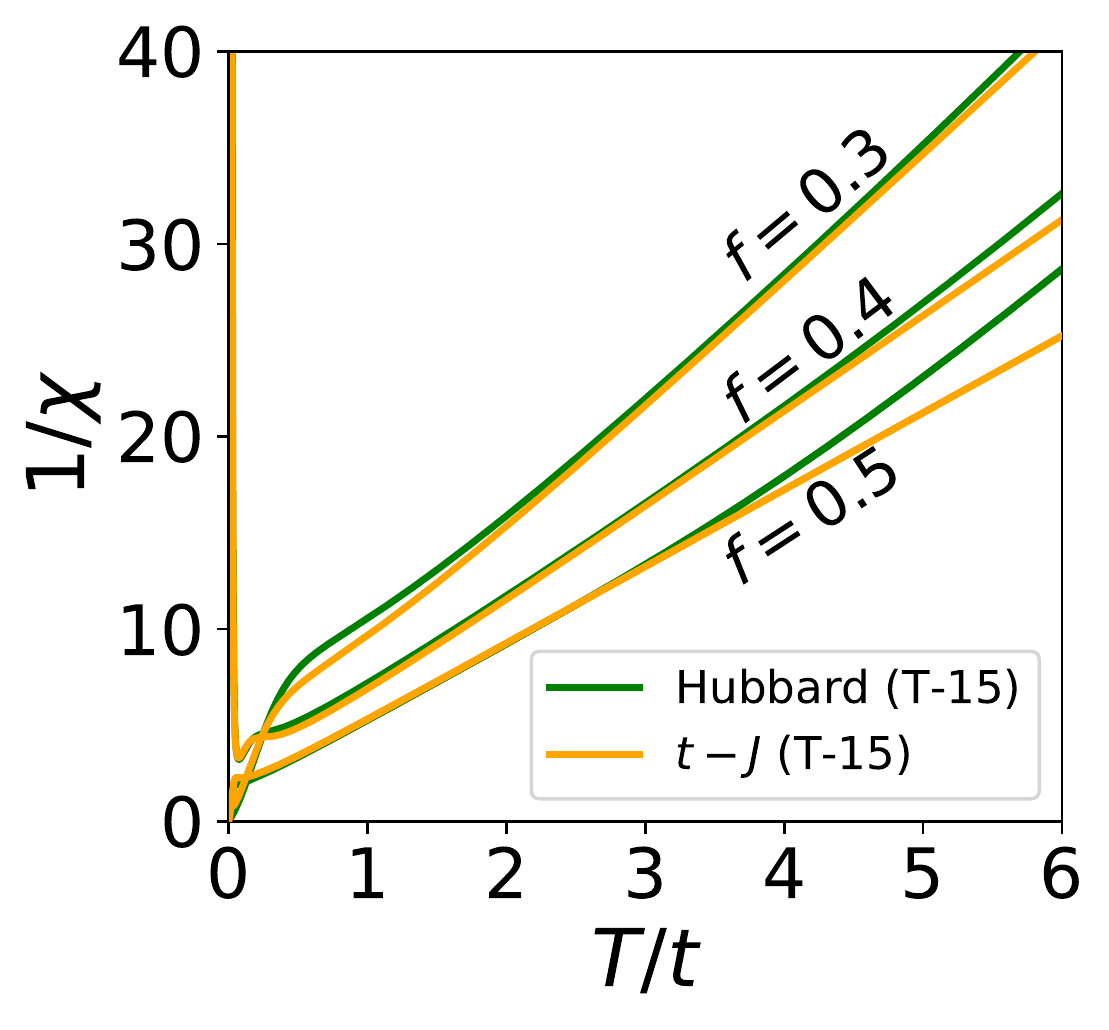}
\subfigimg[height=106pt]{\qquad\qquad\qquad\qquad;\;\textsf{\normalsize(b)}}
{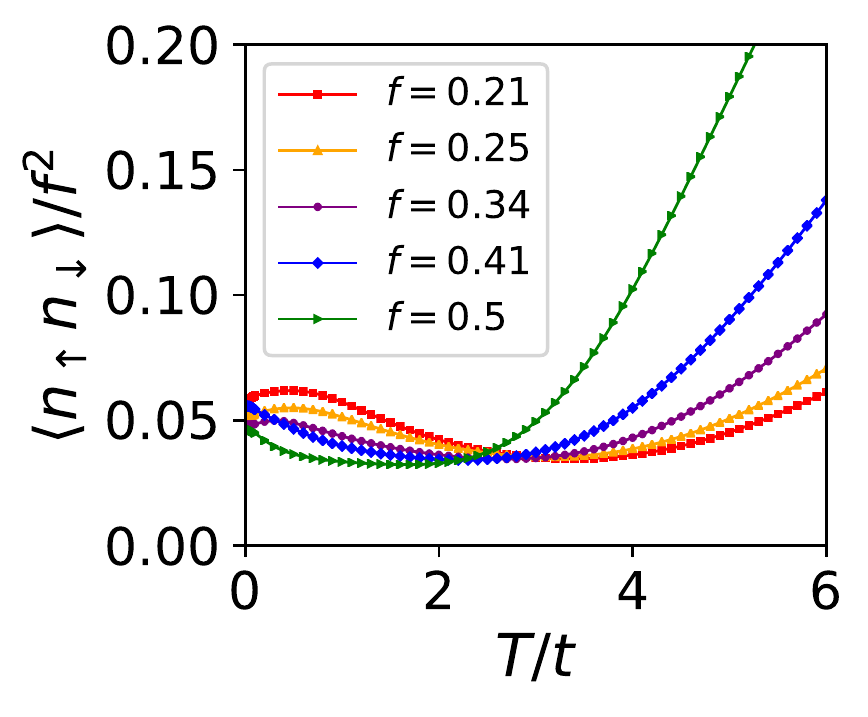}
\caption{
(a) Inverse susceptibility ($1/\chi$) versus temperature ($T$, in units of $t$) for the T-15 cluster for representative $f$ for both the Hubbard 
and $t-J$ models using $U/t=20$ and $J/t=0.20$ respectively.
(b) Normalized on-site double occupancy correlator $\langle n_{i,\uparrow} n_{i,\downarrow}\rangle$ for the T-12 Hubbard model using the same parameters as in (a). 
}
\label{fig:Hubbard_tJ_2}
\end{figure}

\section{Comparison of $t-J$ and Hubbard models}
\label{sec:apptJ}
In Section~\ref{sec:tJ} we developed insights based on the $t-J$ model. Here we comment further on the relation between the Hubbard and $t-J$ models. 

The overall susceptibility of the two models, see Fig.~\ref{fig:Hubbard_tJ_2}(a), match in the intermediate and low temperature regimes, 
with expected deviations at higher temperature (a scale which is filling dependent). 
In Fig.~\ref{fig:Hubbard_tJ_2}(b) we identify this scale by monitoring the temperature 
dependence of the on-site double occupancy correlator, $\langle n_{i,\uparrow} n_{i,\downarrow}\rangle$ (for an arbitrary site $i$), 
normalized with respect to its expected value for the non-interacting case ($f^2$) to facilitate comparison between different densities.
At low temperatures, the double occupancy correlator is small across all densities, however there is a shallow (but prominent) dip in its value as the temperature is increased. 
This observation has been recently made elsewhere as well, where it was attributed to a Pomeranchuk effect associated with the high entropy of states at 
intermediate temperatures~\cite{Wietek_PRX}. 

On increasing the temperature further, the double occupancy correlator becomes appreciably large at a temperature 
that is a small fraction of $U$ (i.e., well below $20t$). This temperature is strongly dependent on the filling: the susceptibility for $f=1/2$ deviates from the $t$-$J$ model at lower $T$ compared to the $f<1/2$ case.
This is because at small $f$ the increased phase space for the motion of the holes of opposite spin types means that they can more effectively avoid each other, thereby circumventing the large Hubbard energy cost. This makes the $t$-$J$ approximation valid with respect to the Hubbard model for a larger temperature range. Once doublon (spin-0) formation becomes increasingly entropically favorable at intermediate and high temperatures in the Hubbard model, it leads to a reduction in magnetic susceptibility (i.e, increase in $1/\chi$) with respect to the $t$-$J$ model.

\begin{figure}
\includegraphics[width=\linewidth]{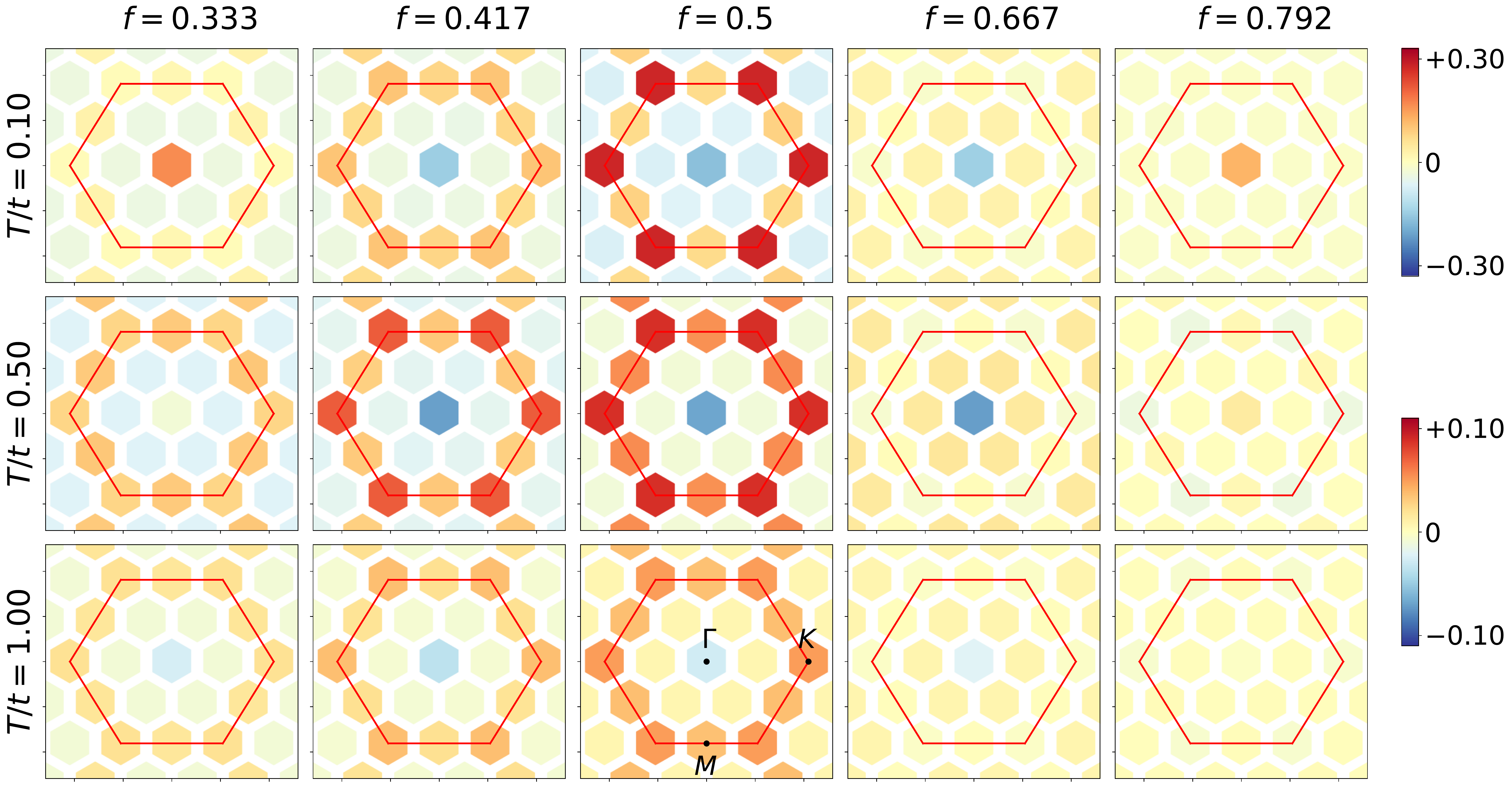}%
\caption{Static structure factor with a high temperature subtraction $S^{zz}(\bfq,T)- S^{zz}(\bfq,5t)$, for the triangular Hubbard model ($U/t=20$) across various fillings $f$ for the T-12 cluster for $T/t=0.1,0.5,1$. The red hexagon in each panel marks the Brillouin zone boundary, and prominent momentum points ($\bf{ \Gamma, K, M}$) are indicated. $M=150$ Krylov vectors were used in the FTLM with $R=500$ random seeds for $f=0.417$ and $f=0.5$ and $R=1500$ for the other fillings.}
\label{fig:T12_ssf}
\end{figure}

\section{Finite Temperature static spin structure factor}
\label{sec:ssf_other}
\para{}
In this Appendix we discuss some aspects of the SSF that facilitate further interpretation of our observations.

For the ${\bf \Gamma}$ point, the SSF is $S^{zz}({\bf q} =  (0,0),T) = \frac{1}{N} \langle S_z^2\rangle_{\mathrm{th}}$, thus for a FM ground state the SSF scales as $N$. 
In case of a FM ground state, multiple $S_z$ sectors are degenerate, i.e., the ED spectrum shows a ground state multiplet with total spin $S \neq 0$. 
Strictly speaking, long-range FM can occur only at $T=0$ since the Hohenberg-Mermin-Wagner theorem rules out true long range order at finite temperature in a two (or lower) dimensional system with continuous symmetry \cite{mermin1966prl,hohenberg1967pr}. 

In Section~\ref{sec:ssf} we presented calculations for the SSF, after subtracting out the high temperature ($T/t=5$) signal, for the nearest-neighbor Hubbard model with $U/t=20$ on the triangular T-15 cluster for various representative fillings. In Fig.~\ref{fig:T12_ssf} we show the analogous calculation for the T-12 cluster. Many qualitative conclusions persist, including (1) the weakening of correlations at the ${\bf K}$ points on doping (i.e. going to lower $f$ starting from $f=1/2$) at low temperature and  (2) the occurrence of FM in the high density regime. Curiously though, a FM ground state was seen at $f=1/3$ which we address in Appendix~\ref{sec:f_1_b_3}.

We motivate the reason for plotting the subtracted SSF.
According to the Curie-Weiss theory, $\chi = \frac{C}{T - \Theta}$. 
Assuming this holds at two temperatures, one ``low" ($T_l$) and one ``high" ($T_h$), and using $\chi T = S^{zz}({\bf q}={\bf \Gamma},T)$,  we get,
\begin{equation}
S^{zz}({\bf \Gamma},T_{l}) (1-\Theta/T_{l}) = S^{zz}({\bf \Gamma},T_h)(1- \Theta/T_h)
\end{equation}
Ignoring the $\Theta/T_h$ term, a reasonable assumption for $\Theta \ll T_h$, we get,
\begin{equation}
	\Theta = \frac{S^{zz}({\bf \Gamma},T_l) - S^{zz}({\bf \Gamma},T_h)}{S^{zz}({\bf \Gamma},T_l)/T_l}
\end{equation}
Thus the subtracted SSF at the ${\bf \Gamma}$ point, but divided by $S^{zz}({\bf \Gamma},T_l)/T_l$, is precisely the CW temperature. 
However it must be noted, as was highlighted in the main text, $\Theta$ itself is temperature-dependent in the intermediate temperature regime 
because $1/\chi$ is not perfectly linear with $T$. 

\begin{figure*}
\subfigimg[height=130pt]{\textsf{\normalsize(a)}}
{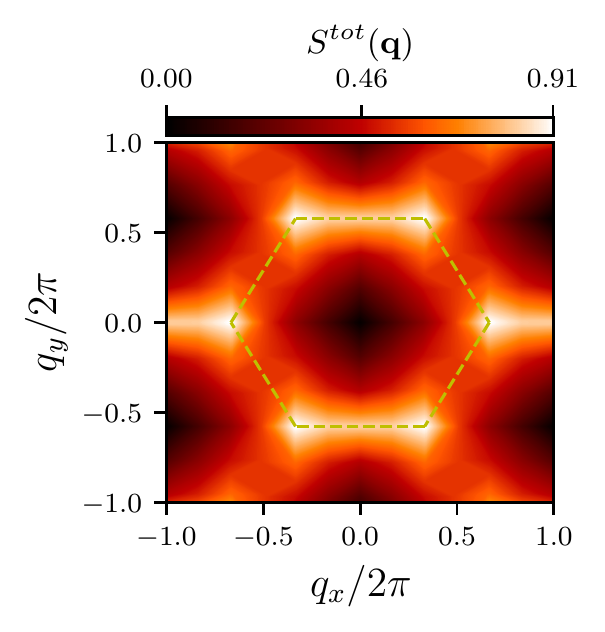}
\subfigimg[height=130pt]{\textsf{\normalsize(b)}}
{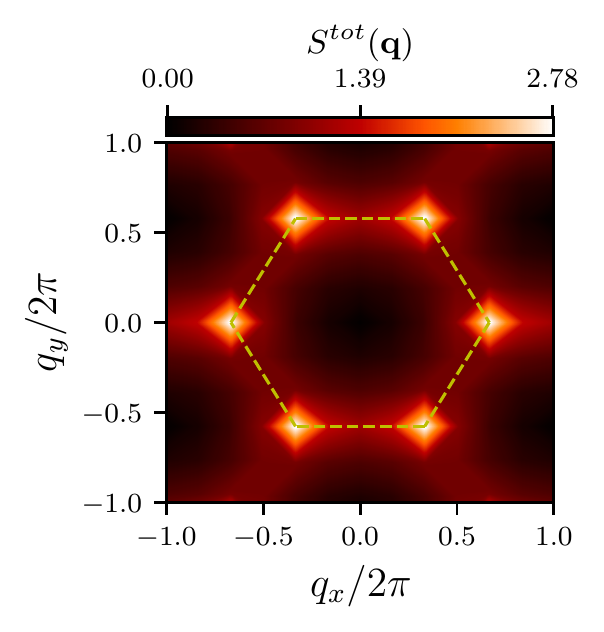}
\subfigimg[height=130pt]{\textsf{\normalsize(c)}}
{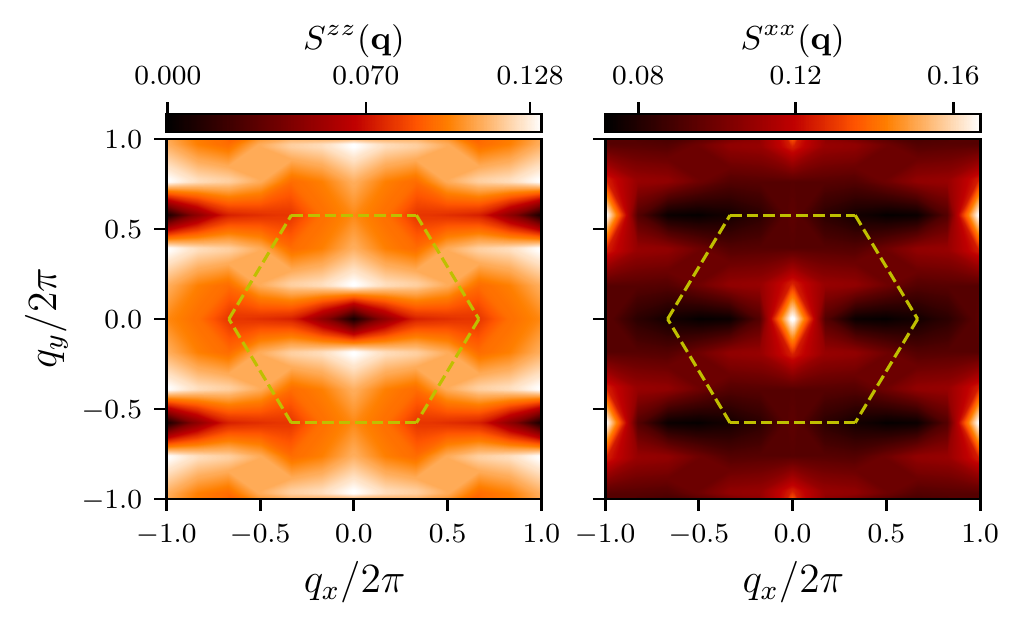}
\caption{
Static spin structure factor from DMRG for the $S_z = 0$ ground state of the length 6 XC-6 cylinder (36 sites). The plots in (a) and (b) show the sum $S^{tot}({\bf q})$ over all three channels ($xx,yy,zz$, which are individually identical) and correspond to fillings (a) $f = 1/3$, 12 up and 12 down electrons and (b) $f = 1/2$, 18 up and 18 down electrons. (c) corresponds to the case of $f \approx 0.806$, 29 up and 29 down electrons, and where the $zz$ and $xx$ ($yy$) channels are shown separately.
The yellow dashed hexagon in each panel marks the Brillouin zone boundary. 
}
\label{fig:DMRG_sf}
\end{figure*}

\begin{figure*}
\subfigimg[height=140pt]{\textsf{\normalsize(a)}}
{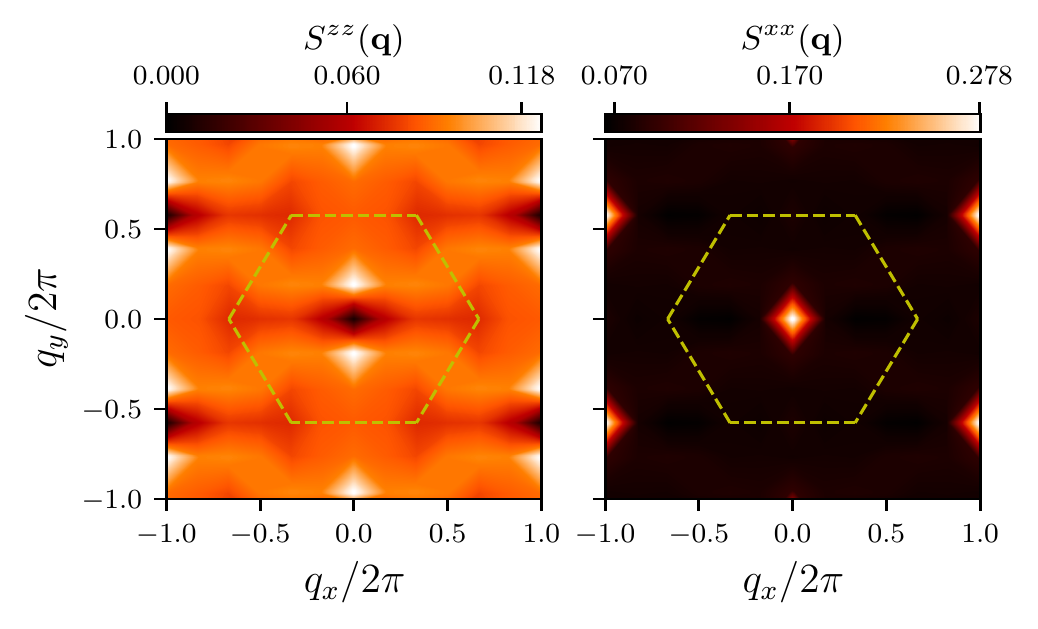}
\subfigimg[height=140pt]{\textsf{\normalsize(b)}}
{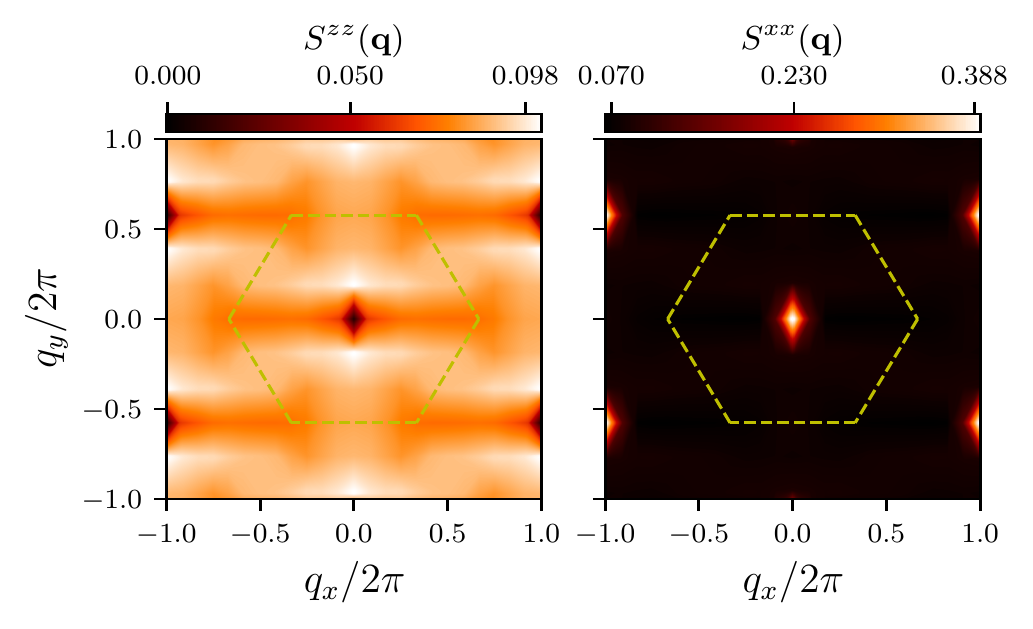}

\caption{
Static spin structure factor from DMRG (for $zz$ and $xx$ channels) at filling $f \approx 0.833$ on the XC-6 cylinder of (a) length 6 (36 sites) and (b) length 12 (72 sites).
}
\label{fig:finiteSizeScale}
\end{figure*}

\section{Ground state DMRG static spin structure factor}
\label{sec:appssf}
Generalizing the SSF to other channels and taking the limit of zero temperature, we have, 
\begin{equation}
	S^{\alpha \alpha}(\bfq) \equiv S^{\alpha \alpha}({\bf q},T \rightarrow 0) \equiv \frac{1}{N} \sum_{i,j} e^{-i\bfq \cdot (\bfr_i - \bfr_j)} \langle \psi_0 | S^{\alpha}_i S^{\alpha}_j |\psi_0 \rangle
\end{equation}
where $\alpha=x,y,z$ and $|\psi_0 \rangle$ is the ground state of the system. (For the case of degenerate states, one must sum over all distinct ground states).
For a rotationally symmetric (singlet) ground state, which is the case for the triangular Hubbard model for many (but not all) fillings,
$S^{zz}(\bfq) = S^{xx}(\bfq) = S^{yy}(\bf q)$. For degenerate ground states, as is the case for a FM, 
choosing a single state from the degenerate multiplet and then computing the expectation values with it does not satisfy this condition.

In the main text we presented results of DMRG calculations on XC-6 cylinders ($36$ and $72$ sites) using a bond dimension of $16000$ targeting the 
ground state in the $S_z = 0$ sector and computed real space spin-spin correlation functions with respect to a reference chosen site. 
In Fig.~\ref{fig:DMRG_sf}, we complement the real space pictures by plotting the SSF for representative cases on length 6 XC-6 cylinders. As expected, for $f=1/2$ the (Bragg) peaks are at the $\bfK$ point of the Brillouin zone, consistent with 120 degree spiral order (Note that the $xx$, $yy$ and $zz$ channels are identical for the singlet ground state and are summed to yield $S^{tot}(\bf q)$). In comparison, the weight at the $\bfK$ points is clearly diminished for $f=1/3$. 
For $f \approx 0.806$, the $xx(yy)$ and $zz$ channels are clearly different. For the $xx (yy)$ channel there is a peak at ${\bf q = \Gamma} $, consistent with FM. 
In the $zz$ channel there is no intensity associated with the ${\bf \Gamma}$ point, this is a consequence of the sum rule corresponding to total $S_z =0$. 

We check for finite size effects to build further confidence in our findings. For example, Fig.~\ref{fig:finiteSizeScale} shows our results for the case of the FM at $f \approx 0.833$ 
on length 6 and 12 XC-6 cylinders. The SSF is visually similar, however, on increasing the size the weight at the ${\bf \Gamma}$ point is found to increase. For length 6, 
$\langle S^2 \rangle$ associated with the ground state is found to be $\approx 20$, (corresponding to $S=4$) and for length 12 it is $\approx 56$ (corresponding to $S=7$). 
This is consistent with a Bragg peak, signalling long-ranged FM, although larger system sizes would be required to establish this definitively.

\begin{figure*}
\includegraphics[width=\linewidth]{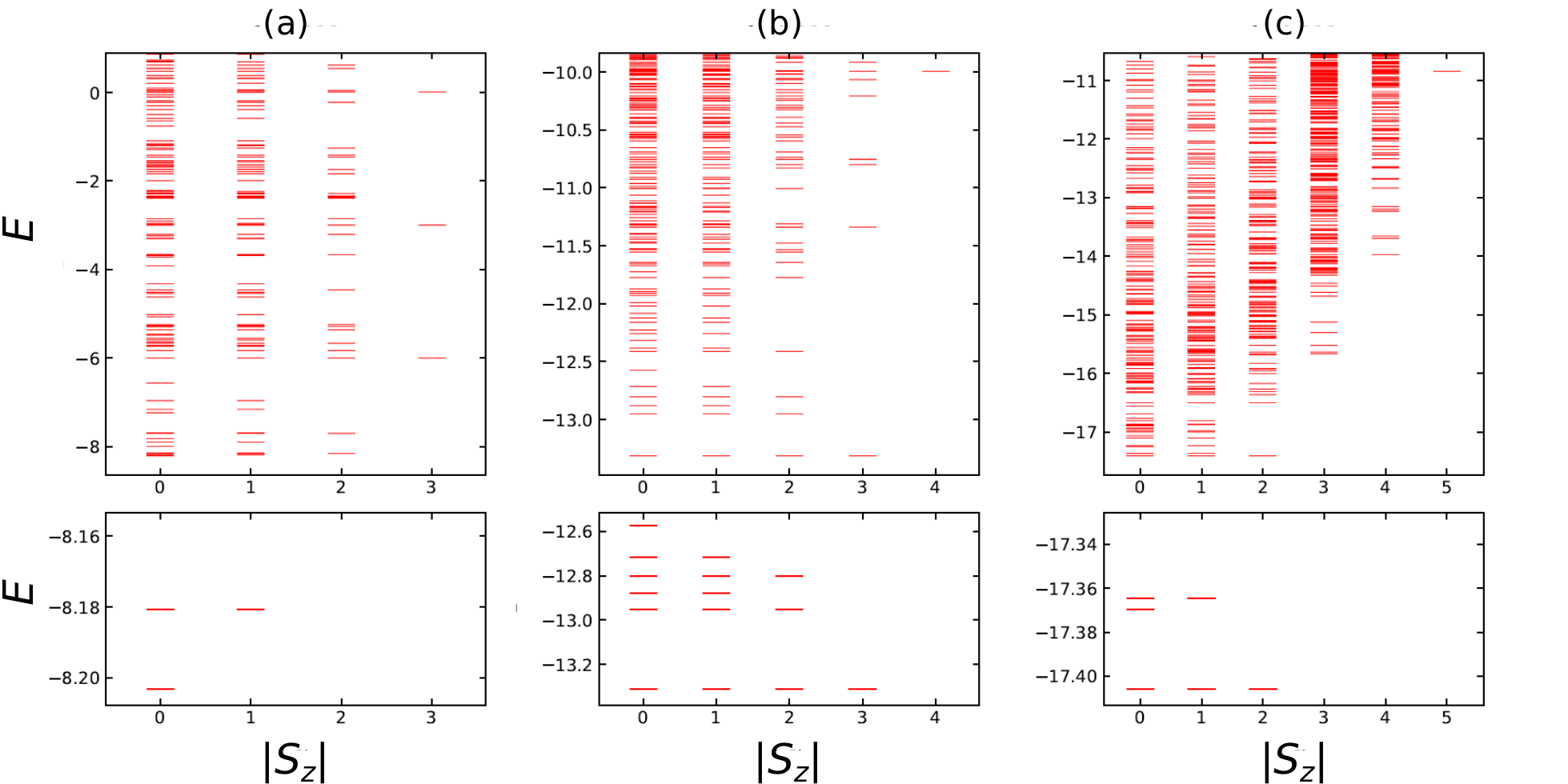}
\caption{Exact diagonalization spectra for the (a) T-9, (b) T-12 and (c) T-15 clusters for $f=1/3$. The lower panels highlight the multiplet structure of the ground state. Note the small scale of the energy gaps, which required further analysis on a bigger system with DMRG.}
\label{fig:f_1b_3_triangular}
\end{figure*}

\section{Ground state for $f=1/3$}
\label{sec:f_1_b_3}
The T-12 and T-15 clusters have a FM ground state for $f=1/3$. Even the T-9 cluster shows a low energy multiplet in close competition with singlets in the spectrum. 
Fig.~\ref{fig:f_1b_3_triangular} shows the gap of the FM to other states decreasing by a factor of $\approx 8$ on going from T-12 to T-15 revealing multiple competing states. 
This required us to further investigate larger clusters with DMRG. Our DMRG results suggest the ground state is not a FM, but one which displays short range AFM correlations, 
which can be seen prominently in Fig.~\ref{fig:S_q}(b).

\bibliography{triangularhubbard}

\end{document}